# Ternary liquid crystalline mixture showing broad antiferroelectric smectic $C_A$* and glassy hexatic smectic $X_A$* phases

Aleksandra Deptuch[1,*], Anna Drzewicz[1], Marcin Piwowarczyk[1], Michał Czerwiński[2], Mateusz Filipow[2], Mateusz Pączek[3], Ewa Juszyńska-Gałązka[1,4]

[1] Institute of Nuclear Physics, Polish Academy of Sciences, Radzikowskiego 152, PL-31342 Kraków, Poland
[2] Institute of Chemistry, Military University of Technology, Kaliskiego 2, PL-00908 Warsaw, Poland
[3] Faculty of Chemistry, Jagiellonian University, Gronostajowa 2, PL-30387 Kraków, Poland
[4] Research Center for Thermal and Entropic Science, Graduate School of Science, Osaka University, 560-0043 Osaka, Japan
*corresponding author, aleksandra.deptuch@ifj.edu.pl

**Abstract**

A ternary liquid crystalline mixture was designed to obtain a tilted hexatic smectic phase in the glassy state. Structural, electro-optic, and dielectric properties of the mixture are investigated, and selected measurements are also performed for its pure components. In particular, the electron density profile perpendicular to smectic layers is determined from the X-ray diffraction data and compared to the results of density functional theory calculations both for the mixture and pure components. Comparison of the experimental smectic layer spacing and tilt angle in the mixture allows us to assess whether molecular dimerization is likely to occur. On the mesoscopic scale, the helical pitch is determined in the $SmC_A$* phase of the mixture, and selective reflection of light is observed under a polarizing microscope in the SmC*, $SmC_A$*, and $SmX_A$* phases. The glass transition in the smectic $X_A$* phase is observed in calorimetric results. At the same time, the dielectric spectra do not directly reveal the primary α-process, although the secondary β- and γ-processes are detected. Overall, the results show that the ternary mixture stabilizes a broad $SmC_A$* phase and enables vitrification of the hexatic $SmX_A$* phase, while the structural data suggest a change in the molecular organization between the SmC* and $SmC_A$* phases.

## 1. Introduction

The glass transition, or vitrification, is associated with a strong slowdown of molecular relaxation, which prevents thermodynamic equilibration, including crystallization [1]. This enables preservation of, e.g., desired optical properties or charge mobility of a less-ordered phase in a given material below the glass transition temperature [2,3]. The glass transition can occur for materials which have any kind of positional, orientational, or conformational disorder, starting from isotropic liquid or amorphous polymer [4-7], via various liquid crystalline phases [7-14] to orientationally-disordered (ODIC) [15] and conformationally-disordered (CONDIS) [16] crystal phases. In this work, we focus on smectic phases, i.e., lamellar liquid crystalline phases formed by elongated molecules, with molecular tilt with respect to the smectic layer normal given by the magnitude $\theta$ and phase $\varphi$. The smectic phases formed by chiral molecules include, in order of decreasing temperature [17-19]: smectic A* (SmA*) – random $\varphi$ and average tilt angle equal to zero; SmC$_\alpha$* – $\varphi$ twisting helically from layer to layer with a short pitch (~ a few smectic layers); SmC* – approximately constant $\varphi$ in neighboring layers, twisting helically with a long pitch (~ hundreds of smectic layers); SmC$_{F2}$* – $\varphi$ twisting with a period of four layers; SmC$_{F1}$* – $\varphi$ twisting with a period of three layers; SmC$_A$* – $\varphi$ changing by approximately 180° from layer to layer, twisting helically with a long pitch (~ hundreds of smectic layers). The magnitude $\theta$ is non-zero in SmC* and all its variants; it can be also non-zero in SmA* (de Vries phase) [18].

In all listed smectic phases, the ordering of molecules within smectic layers is short-range (liquid-like). On decreasing temperature, SmA*, SmC*, and SmC$_A$* phases may transform into their hexatic equivalents, where intra-layer domains with a local hexagonal order are larger than in SmA*, SmC*, SmC$_A$*, and they are oriented in the same direction. SmB$_{hex}$* is a hexatic counterpart of SmA*. For tilted SmC* phase, the hexatic version is either SmF* or SmI* with different orientation of hexagonal domains with respect to the tilt phase $\varphi$ (and similarly SmF$_A$* or SmI$_A$* for SmC$_A$*) [19-24]. SmF* (SmF$_A$*) and SmI* (SmI$_A$*) can generally be distinguished only by X-ray diffraction on the oriented sample [20]. The notation SmX* (or SmX$_A$* for the antiferroelectric analogue) is used throughout this paper.

The chiral tilted smectic phases have potential applications in liquid crystal displays thanks to their ability of bistable or tristable switching in electric field [18,25] and for selective reflection of light (SRL) thanks to a helical ordering [26,27]. The recent results for (*S*)-4′-(1-methylheptyloxycarbonyl)biphenyl-4-yl 4-[5-(2,2,3,3,4,4,4-heptafluorobutoxy)pentyl-1-oxy]-2-fluorobenzoate indicate that a helix pitch in the glassy SmC$_A$* phase is rather constant, which opens the possibility of application in optical filters, although the glass transition temperature of this compound is low, ca. 230 K [28]. The transition to a hexatic phase is reported to change a helix pitch: references [23,24] mention a decrease of a helix pitch and results from [29] indicate an increase of a helix pitch in a hexatic phase. Despite extensive studies on chiral smectic phases, the helical order in the SmX$_A$* phase has received limited attention, especially in the glassy state.

This study focuses on the tertiary mixture of (*S*)-4-[(1-methylheptyloxy)carbonyl]phenyl 4′-octyloxy-4-biphenylcarboxylate (MHPOBC), (*S*)-4'-(1-methylheptylcarbonyl)biphenyl-4-yl 4-[2-(2,2,3,3,4,4,4-heptafluorobutoxy)etyl-1-oxy]benzoate (3F2HPhH6), and (*S*)-4'-(1-methylheptylcarbonyl)biphenyl-4-yl 4-[3-(2,2,3,3,4,4,4-heptafluorobutoxy)propyl-1-oxy]benzoate (3F3HPhH6), with the MHPOBC : 3F2HPhH6 : 3F3HPhH6 molar ratio 0.5 : 0.25 : 0.25. The mixture is further denoted as MIX23HH6. The molecular formulas of the MIX23HH6 components are presented in Figure 1. MHPOBC has the phase sequence Cr (357 K) SmC$_A$* (391 K) SmC$_{FII}$* (392 K) SmC* (394 K) SmC$_\alpha$* (395 K) SmA* (422 K) Iso on heating (the exact narrow temperature ranges of SmC$_{FII}$*, SmC*, SmC$_\alpha$* phases vary slightly between sources) [30,31]. Additionally, a monotropic

SmX$_A$* phase is formed at 338 K on the cooling run only [30,32]. The glass transition of SmX$_A$* of MHPOBC is not observed for the cooling rates up to 40 K/min [32]. The phase sequence of 3F2HPhH6 on heating is Cr (342 K) SmC$_A$* (388 K) SmC* (401 K) Iso and that of 3F3HPhH6 is Cr (352 K) SmC$_A$* (389 K) Iso [27]. The presence of monotropic phases and glassforming properties of these two compounds have not been investigated yet, but the results for longer 3FmHPhH6 homologs with m = 4, 5, 7 indicate that the hexatic SmX$_A$* phase may be present on supercooling [33,34]. Therefore, the formation of SmX$_A$* in the mixture can be expected.

The main aims of this work are to provide new data on SRL in a hexatic phase and its glassy state, and to compare the structure of smectic phases in the mixture with those in pure components. In the first part, the behavior of supercooled pure 3F2HPhH6 and 3F3HPhH6 components is investigated together with the MIX23HH6 mixture by the differential scanning calorimetry (DSC) and polarizing optical microscopy (POM) methods at selected cooling/heating rates. In the second part, the structure of smectic phases of MIX23HH6 and its components is investigated by the X-ray diffraction (XRD). The smectic layer spacing is determined from positions of low-angle diffraction peaks and electron density distribution perpendicular to smectic layers is obtained from the integrated intensities of these peaks. In the third part, the dielectric strength and relaxation times in 3F2HPhH6, 3F3HPhH6, and MIX23HH6 are determined by the broadband dielectric spectroscopy (BDS). The corresponding results for MHPOBC were reported elsewhere [32], therefore, only extended XRD data of this component are presented (results in [32] did not include electron density profile). The fourth and fifth parts focus exclusively on MIX23HH6. They cover SRL originating from the helical order as well as its electrical, electro-optical, and switching behavior in an electric field (specifically spontaneous polarization, tilt angle, and switching time).

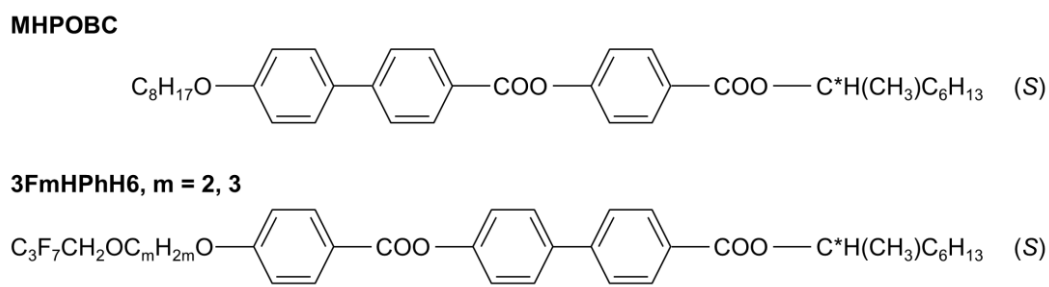

Figure 1. Molecular structures of the MHPOBC and 3FmHPhH6 (m = 2, 3) compounds.

## 2. Experimental and computational details

The (S)-4-[(1-methylheptyloxy)carbonyl]phenyl 4′-octyloxy-4-biphenylcarboxylate (MHPOBC), (S)-4'-(1-methylheptylcarbonyl)biphenyl-4-yl 4-[2-(2,2,3,3,4,4,4-heptafluorobutoxy)etyl-1-oxy]benzoate (3F2HPhH6), and (S)-4'-(1-methylheptylcarbonyl)biphenyl-4-yl 4-[3-(2,2,3,3,4,4,4-heptafluorobutoxy)propyl-1-oxy]benzoate (3F3HPhH6) were synthesized in the Institute of Chemistry of the Military University of Technology in Warsaw according to synthetic routes described in [35] (MHPOBC) and [27,36] (3F2HPhH6, 3F3HPhH6). The ternary mixture MIX23HH6 of MHPOBC : 3F2HPhH6 : 3F3HPhH6 with the molar ratio 0.5004(4) : 0.2491(3) : 0.2505(3) was prepared by dissolution of pure compounds in acetone. After evaporation of acetone, the mixture was heated to 433 K to the isotropic liquid phase and cooled down to room temperature.

The DSC measurements were performed under a nitrogen atmosphere for the 3F2HPhH6, 3F3HPhH6, and MIX23HH6 samples weighing 6.69, 8.41, and 5.75 mg, respectively, and contained within aluminum pans. The DSC scans were collected with DSC 2500 (TA Instruments) calorimeter at 2, 5, 10, 15, 20, 25, 30, 35, and 40 K/min cooling/heating rates in the temperature range between 173 K and 433 K. First, each sample was heated to 433 K at 10 K/min (1st heating) and then, the measurements for each rate were performed in the cooling-heating cycle. The DSC results were analyzed with TRIOS software.

The POM observations in the transmission mode were performed for the 3F2HPhH6, 3F3HPhH6, and MIX23HH6 films placed between two thin glass slides without any alignment layers. The cooling/heating rate was 10 K/min for all samples, and additionally 40 K/min for 3F2HPhH6 and 1 K/min for 3F3HPhH6. The observations in the reflection mode were conducted for MIX23HH6 in an electro-optic cell with 5 μm thickness and providing a homeotropic alignment. The POM textures were captured using Leica DM2700 P polarizing microscope and Linkam temperature stage. The weighted average intensity of each image was calculated using TOApy [37], while red, green, and blue channel components were determined using ImageJ [38].

The XRD measurements were performed for the MHPOBC, 3F2HPhH6, 3F3HPhH6, and MIX23HH6 flat samples in a sample holder of dimensions 13 mm × 10 mm × 0.2 mm. The XRD patterns were collected with the X'Pert PRO (PANalytical) diffractometer in the Bragg-Brentano geometry, using CuKα radiation ($\lambda_{CuK\alpha 1}$ = 1.540562 Å, $\lambda_{CuK\alpha 2}$ = 1.544390 Å [39]), on cooling from 433 K to 298 K. The calibration of the 2θ position was done using the NIST Standard Reference Material 675 [40], supplied by Merck. To compensate for small changes in the sample height after the first heating run, the additional systematic shift in the 2θ angle was corrected based on the positions of the 2nd, 3rd, and 4th diffraction peaks deep in the SmC$_A$* or SmX$_A$* phase. The XRD patterns were analyzed in WinPLOTR [41] and OriginPro.

The BDS measurements were carried out for the 3F2HPhH6, 3F3HPhH6, and MIX23HH6 samples of ca. 50 μm thickness placed between gold electrodes with polytetrafluoroethylene spacers. The BDS spectra were collected with the Novocontrol impedance analyzer for frequencies between 0.1 Hz and 10 MHz on cooling from isotropic liquid to 173 K and then on heating back to isotropic liquid. The BDS spectra were analyzed in OriginPro.

The UV-Vis-NIR spectroscopy measurements were performed for the MIX23HH6 sample in a homeotropic alignment with the Shimadzu spectrometer in the range of 360-3000 nm. The helix pitch was determined as $p_h = \lambda_{min}/1.5$ [27,42], where $\lambda_{min}$ is the wavelength of SRL maximum.

The electric field switching studies were performed for the MIX23HH6 sample in a surface-stabilized bookshelf geometry using special driving scheme for the SmC$_A$* phase [43]. The setup comprised the R&S HMF2550 function generator, FLC F20AD amplifier, and Biolar PI – PZO Poland polarizing microscope with a Linkam THMS 600 temperature stage. The spontaneous polarization and tilt angle were determined by the reversal current method [44] with the R&S HM0724 oscilloscope and by observation of the Clark-Lagerwall effect [45] with the Thorlabs PDA100A photodetector, respectively.

The DFT geometry optimizations for isolated molecules were carried out in Gaussian 16, Revision C.01 [46] (B3LYP-D3(BJ) exchange-correlation functional [47-49], 6-31+Gd basis set [50]). The molecules were visualized in Avogadro [51]. The starting model for MHPOBC was based on the experimental molecular conformation in a crystal phase [52,53] and tilt angle from [54]. The starting models for 3F2HPhH6 and 3F3HPhH6 were based on a strongly twisted conformation inferred previously for 3F7HPhH6, which belongs to the same homologous series with an identical fluorinated chain and chiral center, contributing to the molecular shape [34].

**3. Results and discussion**

*3.1. Phase sequence*

The DSC thermograms of each sample are presented in Figure 2. The phase transition temperatures were determined as the onset temperatures [55]. An exception was the pair of overlapping anomalies observed for 3F3HPhH6 at 2 K/min, where the peak temperature [55] of the second anomaly was used. The glass transition temperature was assumed to be at the half-height of the step-like anomaly [56]. The results are presented in Figure 3 as a function of the cooling/heating rate. Phase transition temperatures and corresponding enthalpy changes are presented in Table 1. Complementary POM data are provided in Figures S1-S10 of the Supplementary Materials.

Two smectic phases are observed for 3F2HPhH6 by DSC (Figures 2a, 3a) and POM (Figures S1-S4) above the melting temperature, in agreement with SmC* and SmC$_A$* reported in [27]. 3F2HPhH6 crystallizes on cooling at all applied cooling rates. During fast cooling at 25-40 K/min, there is a distortion of the exothermic anomaly related to crystallization. Four scenarios may account for this distortion: (1) two-step crystallization, (2) crystallization followed by the glass transition of the remaining SmC$_A$* fraction, (3) SmC$_A$* → SmX$_A$* transition followed by crystallization, (4) SmC$_A$* → SmX$_A$* transition followed by the glass transition of SmX$_A$*. The POM textures collected during cooling at 40 K/min (Figure S3) indicate the (3) scenario. The abrupt change of textures' color from pink to green, with a peak in the green component observed at 297 K, is interpreted as the SmC$_A$* → SmX$_A$* transition. The next change in color from green to red is attributed to incomplete crystallization, where numerous small crystallites do not distort the fan-shaped texture. During heating at 15-40 K/min, a broad exothermic anomaly with the onset at 292-305 K is observed in DSC thermograms below the melting temperature of a crystal phase, corresponding to the cold crystallization of the smectic phase supercooled in the prior fast cooling run. The POM observations at the 40 K/min heating rate (Figure S4) confirm cold crystallization; the texture changes gradually in the 300-320 K range from the red fan-shaped one to the brown texture of a crystal phase observed for slower cooling/heating at 10 K/min (Figures S1 and S2).

DSC results at 5-40 K/min (Figures 2b, 3b) and POM observations at 10 K/min (Figures S5 and S6) indicate one smectic phase of 3F3HPhH6, which corresponds to SmC$_A$* reported in [27]. However, DSC results at 2 K/min show the splitting of the anomaly at ~390 K, which suggests an additional smectic phase between SmC$_A$* and isotropic liquid (inset in Figure 2b). On the other hand, POM textures collected at a low 1 K/min rate do not reveal any additional smectic phase (Figures S7 and S8). 3F3HPhH6 crystallizes at all applied cooling rates. During heating, the broad anomaly related to cold crystallization is observed below the melting temperature, but the corresponding enthalpy change is very small, about 1 kJ/mol in total, which means that crystallization is almost complete in the cooling run. We propose that supercooled SmC$_A$*, present in a small fraction below the melt crystallization temperature, undergoes transition to SmX$_A$*; it corresponds to a small anomaly observed at 290-300 K during cooling.

The MIX23HH6 mixture shows four smectic phases, detected both by DSC (Figures 2c, 3c) and POM (Figures S9 and S10): enantiotropic SmA*, SmC*, SmC$_A$*, and monotropic, metastable hexatic SmX$_A$*, which is present below the melting temperature. SmX$_A$* undergoes the glass transition at $T_g$ = 265.5 K on cooling and 273.4 K on heating. Melt crystallization is not observed on cooling, while cold crystallization occurs on heating above $T_g$, with the onset at 304-305 K. The crystal phase melts at 313-320 K and the melting enthalpy decreases from ~10 K/min at 2 K/min to only ~1 kJ/mol at 40 K/min. This explains why the POM texture in the region of cold crystallization observed at 10 K/min (Figure S10, 308 K) is a fan-shaped texture of a smectic phase: cold crystallization is incomplete, and formed crystallites are small.

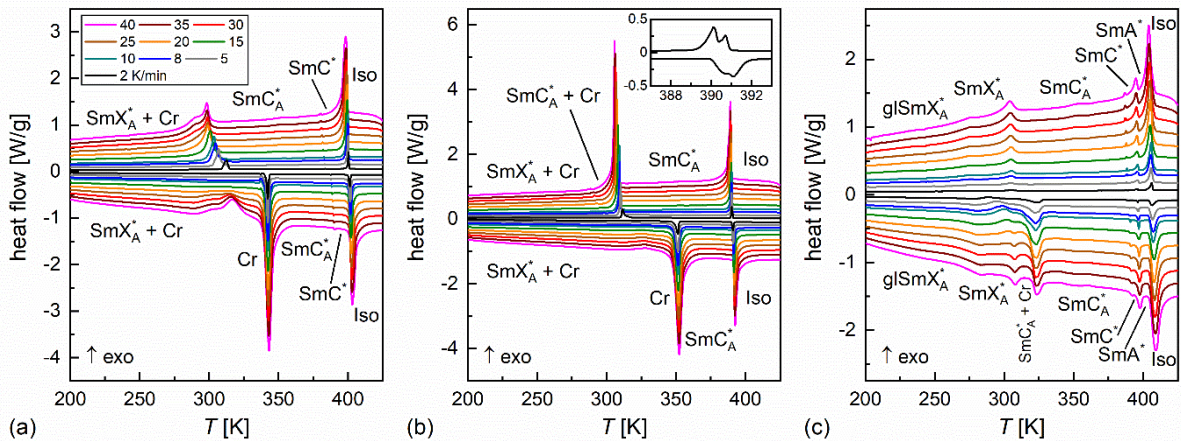

Figure 2. DSC thermograms of 3F2HPhH6 (a), 3F3HPhH6 (b), and MIX23HH6 (c). The legend, common to all panels, shows the cooling/heating rates in K/min. The inset in (b) shows a close-up to the narrow temperature range of a possible additional smectic phase, observed only at 2 K/min.

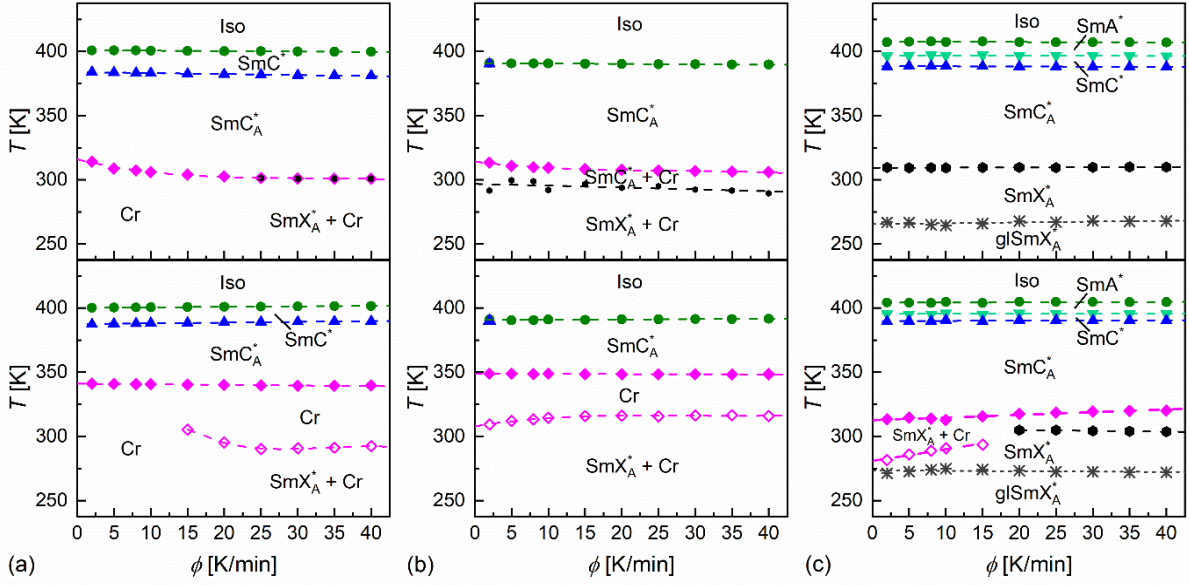

Figure 3. Phase transition temperatures of 3F2HPhH6 (a), 3F3HPhH6 (b), and MIX23HH6 (c) at different cooling rates (upper panels) and heating rates (bottom panels).

Table 1. Phase transition temperatures (upper row, in K, extrapolated to 0 K/min) and enthalpy changes (*bottom row*, in kJ/mol) determined by DSC for 3F2HPhH6, 3F3HPhH6, and MIX23HH6. The presence and absence of a phase is denoted by • and -, respectively.

| sample | Cr | | SmX$_A$* | | SmC$_A$* | | SmC* | | SmA* | | Iso |
|---|---|---|---|---|---|---|---|---|---|---|---|
| 3F2HPhH6, cooling | • | 315.9 *12.8* | • | 289.5[a] *1.3* | • | 383.8 *>0.1* | • | | - | | 400.8 *7.7* | • |
| 3F2HPhH6, heating | • | 341.0 *16.9* | • | | • | 387.4 *0.1* | • | | - | | 400.2 *7.6* | • |
| 3F3HPhH6, cooling | • | 314.1 *13.6* | • | 297 *>0.1*[b] | • | 390.1 *2.3* | - | | - | | 390.8 *4.4* | • |
| 3F3HPhH6, heating | • | 348.8 *19.5* | • | | • | 390.0 | - | | - | | 390.6 *6.9* | • |
| MIX23HH6, cooling | - | | • | 309.3 *1.7* | • | 388.5 *0.2* | • | 396.9 *1.1* | • | 407.5 *5.4* | • |
| MIX23HH6, heating | • | 312.6 *9.8* | • | 306.3 *0.6* | • | 389.7 *0.2* | • | 395.3 *1.1* | • | 404.1 *5.6* | • |

[a] observed only at 25-40 K/min, the SmC$_A$* → SmX$_A$* transition temperature given for 40 K/min
[b] enthalpy change significantly underestimated due to partial crystallization

The isoconversional method was used to determine the effective activation energy $E_{eff}$ of crystallization of pure 3F2HPhH6 (Figure S11) and 3F3HPhH6 (Figure S12). The activation plot in this method is based on the logarithm of conversion rate: $dx/dt$, at a selected conversion degree $x$, plotted against the temperature $T_x$ corresponding to a selected $x$ value [57-60]:

$$\frac{dx(t)}{dt} = f(x) A \exp\left(-\frac{E_{eff}}{RT_x}\right), \quad (1)$$

where $f(x)$ and $A$ are fitting parameters.

The $dx/dt$ and $T_x$ values were determined from DSC thermograms in the region of melt crystallization at different cooling rates: 2-20 K/min for 3F2HPhH6 and 2-40 K/min for 3F3HPhH6. The high cooling rates were excluded from analysis for 3F2HPhH6 because of the SmC$_A$* → SmX$_A$* transition overlapping with crystallization. The $E_{eff}$ values are obtained as a function of $x$, but they can be plotted approximately as a function of temperature by taking middle $T_x$ from each linear part of the $\ln(dx/dt)$ vs. $1000/T_x$ plot [58,59]. Positive $E_{eff}$ indicates that crystallization occurs in a temperature region of high nucleation rate and low diffusion (crystal growth) rate, thus, the overall crystallization kinetics is constrained by diffusion. Negative $E_{eff}$ describes a reverse situation – crystallization occurs in a temperature region of low nucleation rate and high diffusion rate, and the constrain on the overall crystallization is imposed by nucleation [5,61]. Typically, positive $E_{eff}$ is obtained at lower temperatures and negative $E_{eff}$ at higher temperatures; $E_{eff} \approx 0$ characterizes the temperature region of the fastest crystallization [58]. 3F2HPhH6 has two regions of $E_{eff} \approx 0$: at 298-299 K and around 311 K (Figure S11c). In intermediate temperatures, $E_{eff} < 0$, which means nucleation-constrained crystallization. Crystallization of 3F3HPhH6 is the fastest around 307 K and strongly nucleation-constrained around 310 K (Figure S12c). The temperature region of melt crystallization is narrower for 3F3HPhH6 than for 3F2HPhH6.

*3.2. Structure of smectic phases*

The X-ray diffraction patterns of 3F2HPhH6, 3F3HPhH6, and MHPOBC show that all pure components crystallize on slow cooling (Figure 4a). In contrast, MIX23HH6 does not show crystallization on slow cooling down to room temperature (Figure 4b). The low-angle peaks with Miller indices $(00l)$, indicated for MIX23HH6 in Figure 4b, are related to the positional layer order. The layer spacing $d$ is determined from the Bragg equation $l\lambda_{CuK\alpha 1} = 2d \sin\theta_{00l}$ [20,62], where $\theta_{00l}$ is the low-angle $(00l)$ peak position and $l$ is an integer (Figure 4c). MHPOBC shows a layer shrinkage of 3.4% between SmA* and SmC$_A$* and a 6.5% increase in the layer spacing between SmC$_A$* and SmX$_A$*. The layer shrinkage in 3F2HPhH6 and 3F3HPhH6 is small, only 2.4% and 1.9%, respectively. The layer spacing of 3F3HPhH6 increases weakly just below the Iso → SmC$_A$* transition, from 29.6 Å at 392 K to 29.7 Å at 393 K, which corresponds to a double transition observed in DSC thermograms. The layer shrinkage in MIX23HH6 between SmA* and SmC$_A$* is 8.9% and an increase in the layer spacing between SmC$_A$* and SmX$_A$* is 4.0%.

Intensities of $(00l)$ peaks are related to distribution of electron density $\rho(z)$ in a direction perpendicular to smectic layers [21,63,64]. Molecules in smectic phases rotate around their short axes [65]. Consequently, $\rho(z)$ is centrosymmetric with respect to the middle of a smectic layer and is described by series of cosines [64]:

$$\rho(z) = \rho_0 + \sum_{l=1}^{4} \pm |F_{00l}| \cos(2\pi l z/d). \qquad (2)$$

Structure factors $F_{00l}$ are related to integrated intensities of $(00l)$ peaks as $|F_{00l}| = \sqrt{I_{00l}/Lp}$, where the Lorentz-polarization correction for an aligned sample is calculated as [62]:

$$Lp \propto \frac{1+\cos^2(2\theta)}{\sin(2\theta)}. \qquad (3)$$

Intensities of $(00l)$ peaks from smectic layers are much stronger than the diffuse maximum at higher $2\theta$ angles (originating from an intra-layer order [20]) in the XRD patterns of MIX23HH6 and its components. It indicates a mainly homeotropic alignment of samples and Equation (3) for the $Lp$ correction can be applied in this case. The absolute values of $F_{00l}$ obtained from experimental intensities are presented in Figure 5. The largest structure factor for 3F2HPhH6 is $|F_{003}|$, which means that the $\rho(z)$ distribution is mainly described by the $\cos(6\pi z/d)$ component. For 3F3HPhH6, MHPOBC, and MIX23HH6, the largest structure factor is $|F_{001}|$, so the main component of $\rho(z)$ is $\cos(2\pi z/d)$. The signs of $F_{00l}$ are unknown, but the negative (−) sign of the main $F_{00l}$ factor ($F_{001}$ or $F_{003}$) can be inferred when one assumes that $\rho(z)$ has local minima between smectic layers ($z = 0$ and $z = d$) and local maximum around the middle of a smectic layer ($z = d/2$). The signs of minor $F_{00l}$ are more arbitrary because they affect total $\rho(z)$ to a smaller extent. After testing various signs of $F_{00l}$ and comparing with results of DFT calculations (discussed later), the final set used to calculate $\rho(z)$ from Equation (2) is $-|F_{00l}|$ for all $l$, except $+|F_{005}|$ in MIX23HH6 (Figure 6). The electron density profiles of 3F2HPhH6 (Figure 6a) and 3F3HPhH6 (Figure 6b) have local maxima not only around the middle of smectic layers, related to the aromatic core, but also close to the borders of smectic layers, related to the fluorinated chain. These side maxima are particularly strong for 3F2HPhH6, probably because of a closer proximity of a fluorinated chain and the oxygen atoms adjacent to the aromatic ring than for 3F3HPhH6 (Figure 1). The side maxima in $\rho(z)$ are absent in MHPOBC (Figure 6c). The electron density profile in MIX23HH6 resembles that of MHPOBC at higher temperatures, while the side maxima in $\rho(z)$ arise upon approaching the SmC$_A$* → SmX$_A$* transition (Figure 6d).

The $\rho(z)$ profiles were predicted based on DFT calculations. This is not straightforward, as considered compounds have flexible molecules. To reproduce it to some extent, four conformations of similar length and shape were optimized for each compound. The molecular models were properly oriented, so that the z-axis corresponded to the smectic layer normal. The centrosymmetric $\rho(z)$ profiles were obtained by flipping calculated electron density along the center of a smectic layer and averaging with its mirror image. Then they were smoothed by fitting Equation (2) (Figures S13-S15, see also Ref. [28]). Results vary between conformations, therefore, $\rho(z)$ profiles averaged over conformations were calculated (Figure 7). The DFT results for 3F2HPhH6 and 3F3HPhH6 (Figure 7a,b), although do not reproduce quantitatively relative magnitudes of $|F_{00l}|$, show distinct lateral maxima in qualitative agreement with experimental $\rho(z)$ profiles (Figure 6a,b). It is different for MHPOBC, where DFT results predict strong lateral maxima in $\rho(z)$ (Figure 6c), which are absent or weak in experimental $\rho(z)$ (Figure 7c). It can be explained by low smectic order parameters $\tau_l$ for $l > 1$ in MHPOBC, as obtained in Ref. [66] ($\tau_1 \approx 0.8$, $\tau_2 \approx 0.35$, and $\tau_3 < 0.1$), which lowers corresponding intensities of $(00l)$ peaks.

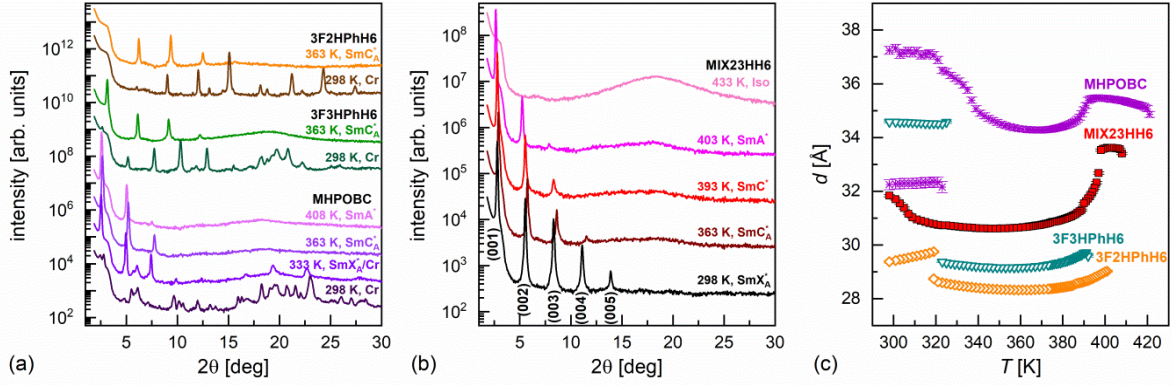

Figure 4. Selected XRD patterns of MHPOBC, 3F2HPhH6, 3F3HPhH6 (a) and MIX23HH6 (b) with the corresponding layer spacing for all samples (c).

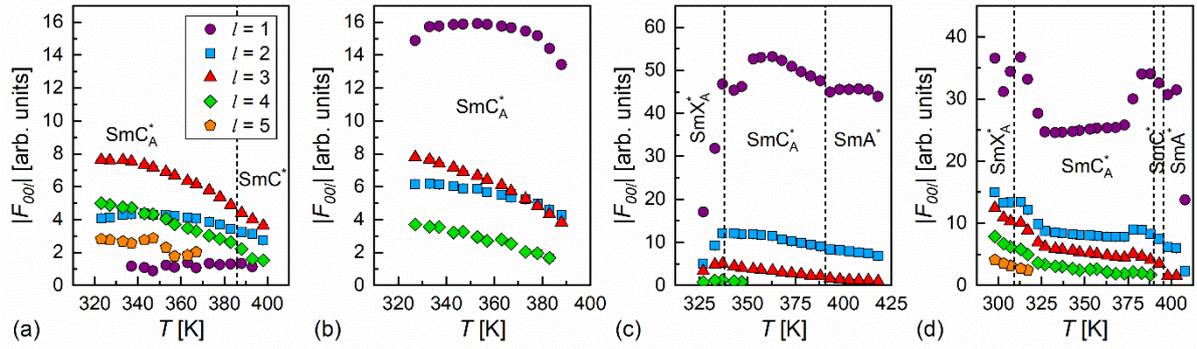

Figure 5. Absolute values of structure factors $|F_{00l}|$ in 3F2HPhH6 (a), 3F3HPhH6 (b), MHPOBC (c), and MIX23HH6. The legend in (a) is common for all panels.

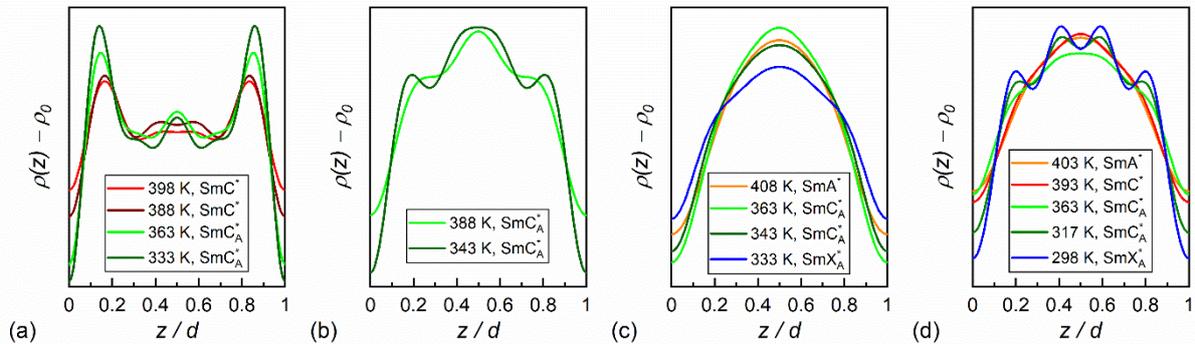

Figure 6. Electron density profiles along the smectic layer normal in 3F2HPhH6 (a), 3F3HPhH6 (b), MHPOBC (c), and MIX23HH6 (d), obtained from XRD patterns.

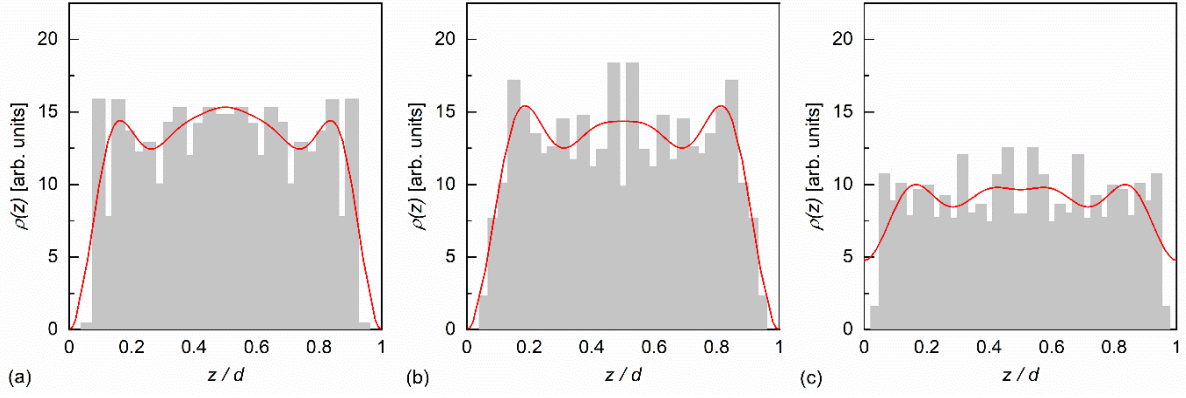

Figure 7. Electron density profiles along the smectic layer normal in 3F2HPhH6 (a), 3F3HPhH6 (b), and MHPOBC (c), obtained from DFT calculations at the B3LYP-D3(BJ)/6-31+Gd level (average over four conformations). Lines are fitting results of Equation (2).

### *3.3. Dielectric relaxation processes*

Representative BDS spectra in smectic phases are shown for MIX23HH6 in Figure 8. The weak soft mode (SM) and strong Goldstone mode (GM) [19,67] are observed in the SmA* phase of MIX23HH6 and SmC* phase of 3F3HPhH6 and MIX23HH6, respectively. The low-frequency $P_L$ phason and high-frequency $P_H$ phason [65,68] are observed in the SmC$_A$* phase of all samples. There is an additional Maxwell-Wagner-Sillars (MWS) process [69] at low frequencies in the SmC* phase of MIX23HH6 and SmC$_A$* phase of 3F3HPhH6 and MIX23HH6. In the hexatic SmX$_A$* phase of MIX23HH6, a new $P_{hex}$ phason [19] appears at the low-frequency side of $P_H$. The α-relaxation expected for glassforming materials [4,7,10,12] is not explicitly visible for MIX23HH6. Finally, in the SmX$_A$* glass of MIX23HH6, the weak secondary β-relaxation and γ-relaxation are observed. There are also weak relaxation processes in the crystal phases of 3F2HPhH6 and 3F3HPhH6, denoted as cr-I, cr-II, cr-III, and cr-IV in an order of increasing frequency (Figure 9). The cr-II process is observed for both compounds, cr-I and cr-IV only for 3F2HPhH6, and cr-III only for 3F3HPhH6.

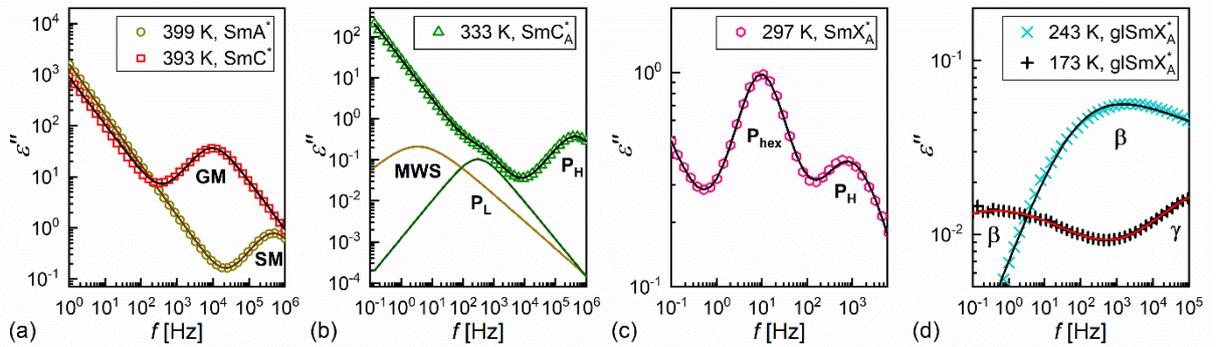

Figure 8. BDS spectra (imaginary part, points) of MIX23HH6 in the SmA* and SmC* phases (a), SmC$_A$* phase (b), hexatic SmX$_A$* phase (c), and SmX$_A$* glass (d). Lines show the fitting results of CC or HN models.

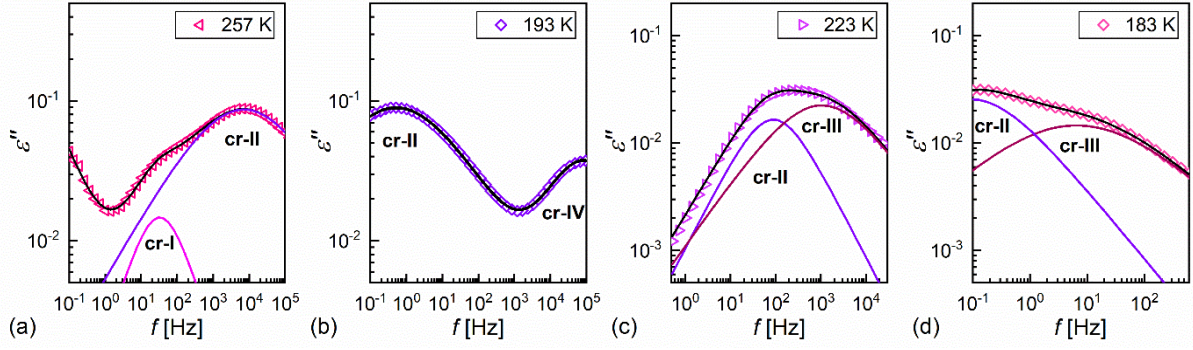

Figure 9. BDS spectra (imaginary part, points) in the crystal phases of 3F2HPhH6 (a,b) and 3F3HPhH6 (c,d). Lines show the fitting results of the CC model.

The relaxation times $\tau$ were determined by fitting to BDS spectra the complex function describing the dielectric permittivity $\varepsilon^*$ based on the Havriliak-Negami (HN) model [19,70,71]:

$$\varepsilon^*(f) = \varepsilon'(f) - i\varepsilon''(f) = \varepsilon_\infty + \sum_j \frac{\Delta\varepsilon_j}{\left(1+(2\pi i f \tau_{HNj})^{1-a_j}\right)^{b_j}} - \frac{iS_1}{(2\pi f)^{n_1}} + \frac{S_2}{(2\pi f)^{n_2}}, \quad (4)$$

where: $\varepsilon_\infty$ is the real part of $\varepsilon^*$ at the frequency limit $f \to \infty$, $\Delta\varepsilon$ is dielectric strength, $\tau_{HN}$ is the HN relaxation time, $a$ and $b$ are shape parameters, $S_1$, $n_1$ and $S_2$, $n_2$ describe the low-frequency background. If $b = 1$, the HN model simplifies to Cole-Cole (CC) model [72] and $\tau_{HN} = \tau$, corresponding to the peak in $\varepsilon''$. If $b \neq 1$, $\tau_{HN}$ does not correspond to the peak in $\varepsilon''$. The relaxation time $\tau$ is then calculated using the formula [71]:

$$\tau = \tau_{HN} \left(\sin\left(\frac{\pi(1-a)}{2+2b}\right)\right)^{-\frac{1}{1-a}} \left(\sin\left(\frac{\pi(1-a)b}{2+2b}\right)\right)^{\frac{1}{1-a}}. \quad (5)$$

Relaxation processes were fitted mainly by the CC model, except the β-process fitted by the NH model. The activation plots of $\tau$ are presented in Figure 10. Most of relaxation processes show an Arrhenius dependence of the relaxation time on temperature, $\tau(T) = \tau_0 \exp(\Delta E/k_B T)$, where: $\tau_0$ is the pre-exponential constant, $\Delta E$ is an activation energy, and $k_B$ is the Boltzmann constant [19]. The $P_L$ process has $\Delta E$ = 119.6(6), 119.7(9), 109.7(7) kJ/mol for 3F2HPhH6, 3F3HPhH6, and MIX23HH6, respectively. The Arrhenius behavior and high $\Delta E$ of $P_L$ correspond to overlapping with the s-process, related to rotations around short molecular axes [62]. The MWS process has $\Delta E$ = 67.3(9) and 68(2) kJ/mol in 3F3HPhH6 and MIX23HH6. The $P_H$ and $P_{hex}$ processes in the SmX$_A$* phase of MIX23HH6 are characterized by $\Delta E$ = 301(4) kJ/mol for $P_H$ and 341(5) kJ/mol for $P_{hex}$. Such large $\Delta E$ values are caused likely by approaching the glass transition. The β-process in MIX23HH6 has $\Delta E$ = 77.0(4) and 22(2) kJ/mol above and below 288 K. The latter value can be interpreted, according to earlier DFT calculations for a very similar molecule [73], as conformational changes within one of terminal chains. The former, higher value includes probably also rotation of the benzene ring. The cr-I process has $\Delta E$ = 68.1(4) kJ/mol. Similar $\Delta E$ values are obtained for cr-II: 60.7(2) kJ/mol for 3F2HPhH6 and 60.3(2) kJ/mol for 3F3HPhH6. The cr-III process shows two $\Delta E$ values, 52.5(3) and 29(1) kJ/mol, above and below 190 K, which may be caused by the transition between two crystal phases. Finally, cr-IV has $\Delta E = 30.5(2)$ kJ/mol. The energy barriers of processes in the crystal phases of 3F2HPhH6 and 3F3HPhH6 indicate their origin in intra-molecular rotations. The energy barrier of 50-60 kJ/mol corresponds to rotation of the benzene ring and that of 30 kJ/mol – to rotation of the chiral chain, according to DFT calculations [73]. It means that the crystal phases of 3F2HPhH6 and 3F3HPhH6 are CONDIS phases. The γ-relaxation time in MIX23HH6 cannot be fitted reliably with the Arrhenius formula as its activation plot is almost horizontal, indicating a very low $\Delta E$.

The $P_H$ process in SmC$_A$* shows a super-Arrhenius behavior described by an empirical Vogel-Fulcher-Tammann (VFT) equation, $\tau(T) = \tau_0 \exp(B/(T - T_V))$, where: $B$ is a constant with a unit of temperature and $T_V$ is the Vogel temperature [19]. The VFT formula is usually applied for fitting of the α-relaxation time and the glass transition temperature $T_g$ is defined as $\tau$ = 100 s [74]. However, if the α-relaxation time cannot be measured, extrapolation of the $P_H$-relaxation time to 100 s enables determination of a slightly overestimated $T_g$ [12]. The expected glass transition temperatures obtained from fitting the VFT formula to the $P_H$-relaxation time in SmC$_A$* are 268(1), 284(2), 276.1(2) K for 3F2HPhH6, 3F3HPhH6, and MIX23HH6. The $P_H$-relaxation time in the SmX$_A$* phase of MIX23HH6 changes its temperature dependence to an Arrhenius behavior and extrapolation to 100 s gives $T_g$ = 268(3) K. Both $T_g$ values obtained from the BDS spectra of MIX23HH6 are close to calorimetric $T_g$ = 265-273 K. The absence of the α-relaxation can be explained by its initial overlapping with the β-relaxation and rapid shifting out of the frequency range at $T_g$, which is rather high compared to $T_g$ = 230-247 K of pure compounds forming the SmX$_A$* glass, where the α-process was observed [33,34,75].

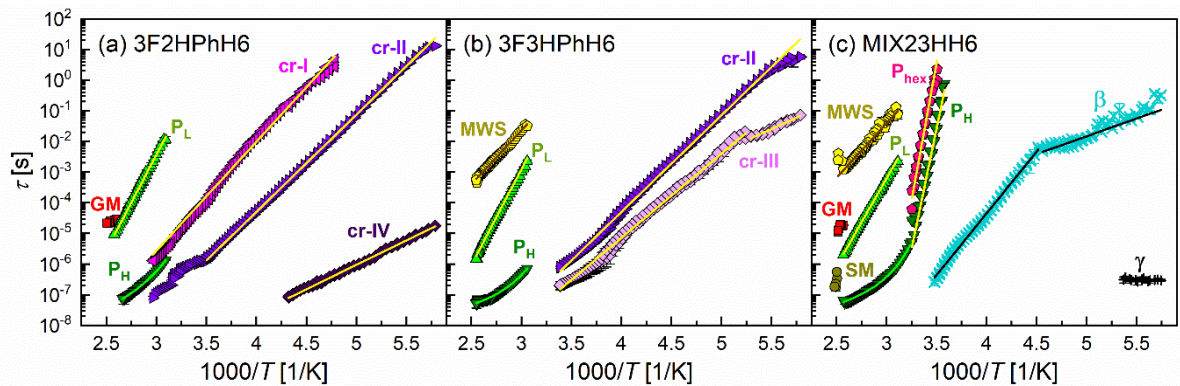

Figure 10. Activation plots of the relaxation times (points) obtained from the dielectric spectra of 3F2HPhH6 (a), 3F3HPhH6 (b), and MIX23HH6 (c) with the fitting results of the Arrhenius or Vogel-Fulcher-Tammann formulas (lines).

## 3.4. Helical order in MIX23HH6

The UV-Vis-NIR spectroscopy measurements in SmC$_A$* show a decrease of a helix pitch $p$ on cooling, with SRL in visible range from 326 K down to the SmC$_A$*/SmX$_A$* transition at 308 K (Figure 11). The helix inversion temperature, estimated by linear extrapolation of $1/p$ to zero, is equal to 356.4(9) K. Another method to observe the helix inversion is POM in the reflection mode, where it shows as a maximum of intensity of all rgb components (Figures S16, S17). The inversion temperature obtained by POM is 360 K on cooling, slightly higher than from spectroscopic results. There is a hysteresis of 19-25 K in the thermochromic effect: maxima of rgb components occur at 318, 314, 310 K on cooling and at 343, 333, 330 K on heating. This shift explains why the inversion is not observed on heating, as it would occur above the transition temperature to SmC*. The POM observations reveal SRL in SmC* of blue light on cooling and blue and green light on heating (Figures S18, S19). The SRL deep in SmX$_A$* is out of the range of a spectrometer, but POM observations show a dark blue image (280 K in Figure S16). It may be a sign of a rather short helix ($p < 200$ nm) in SmX$_A$*. Interestingly, below 260 K there is a color change to dark brown (200 K in Figure S16), which suggests that it has some relationship to the glass transition. However, the exact mechanism of this color change is unknown.

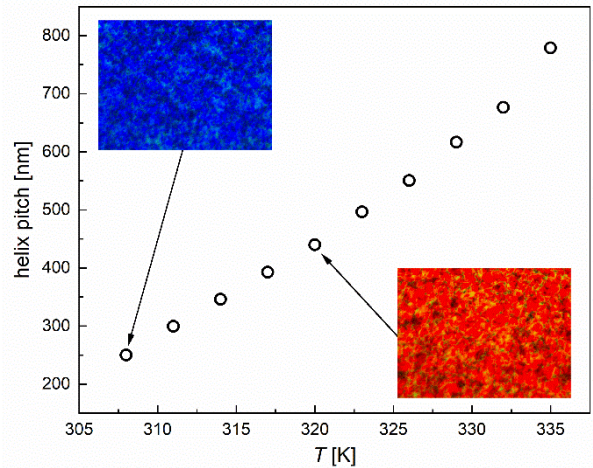

Figure 11. Helix pitch in the SmC$_A$* phase of MIX23HH6 as a function of temperature. POM images collected on cooling in a reflection mode in 320 K and 308 K (622 × 466 μm²) are presented to show the thermochromic effect.

## 3.5. Electro-optical response and switching behavior of MIX23HH6 in an external electric field

The switching time $\tau_{sw}$ in SmC$_A$* increases on cooling from 33 μs in 387 K to 146 μs in 348 K. On further cooling, $\tau_{sw}$ increases more rapidly up to 393 μs at 308 K (Figure 12a). The spontaneous polarization $P_s$ has a maximal value of 258 nC/cm² at 308 K and changes with temperature both in SmC* and SmC$_A$* according to the formula $P_s(t) = P_0(T_c - T)^p$ [76], where $P_0 = 24(4)$ nC/cm², $T_c = 401(1)$ K, $p = 0.50(4)$ (Figure 12b). The tilt angle $\theta$ increases on cooling and reaches 38° close to the SmC$_A$*/SmX$_A$* transition temperature. The non-zero $\Theta$ values of 5-7.5° in SmA* are caused by the electro-clinic effect [19] (Figure 12c). The smectic layer spacing depends on the tilt angle as $d(\Theta) = L \operatorname{acos}(\Theta - \delta\Theta)$, where $L$ is the molecular length and $\delta\Theta$ is a shape parameter, describing deviation of a molecule from a rod-like shape [34]. The experimental $d$ and $\Theta$ values were compared on a scale of a reduced temperature $T - T_c$, where $T_c$ corresponds to the SmA* → SmC* transition, which occurs at a slightly higher temperature in electro-optic measurements than in XRD. The fitting results for MIX23HH6 are $L = 32.31(2)$ Å, $\delta\Theta = 16.7(2)°$ in SmC* and $L = 31.19(5)$ Å, $\delta\Theta = 25.2(5)°$ in SmC$_A$* (Figure 12d). For comparison, mean $L$ and $\delta\Theta$ values obtained by DFT for 3F2HPhH6, 3F3HPhH6, MHPOBC are 29, 30, 32.5 Å and 28°, 25°, 23.5°, respectively. The $L$ value includes non-bonding distance between terminal atoms: 3.22 Å for C, F and 3.59 Å for C, C [77]. The weighted average of calculated $L$ and $\delta\Theta$, based on composition of MIX23HH6, is 31 Å and 25°. The $L$ value is slightly larger and $\delta\Theta$ angle is much lower in SmC*, which can be caused by dimerization of some MHPOBC molecules in a mixture. By dimer we understand two molecules connected by electrostatic interactions, which consequently form one structural unit [21,78]. The dimerization occurs also very likely in pure MHPOBC, as the smectic layer spacing > 34 Å is larger than the length of a single molecule, and it was taken into account while calculating $\rho(z)$ in Figure 7c. Meanwhile, the very good agreement between calculated and experimental $L$ and $\delta\Theta$ values below the SmC* → SmC$_A$* transition suggests that MHPOBC dimers are absent, or at least much less important, in the SmC$_A$* phase of MIX23HH6. This result was applied to predict the $\rho(z)$ profile in MIX23HH6 from DFT calculations. The same molecular models as in Figure 7 and Figures S13-S15 were used, only tilted to match the experimental tilt angle in MIX23HH6 deep in SmC$_A$*. The calculated $\rho(z)$ profile (Figure 13) agrees qualitatively with the experimental one from SmC$_A$* close to the SmC$_A$* → SmX$_A$* transition (Figure 6d). It resembles also the experimental profile in SmX$_A$*, despite the layer spacing increases in the hexatic phase, indicating that dimers, present in SmC* and absent in SmC$_A$*, may form again in SmX$_A$*. The $\rho(z)$ profiles obtained for MIX23HH6 at higher temperatures are closer to the sinusoidal wave than the calculated one, which is explained by low $\tau_l$ order parameters of the layer order, as was also the case for pure MHPOBC.

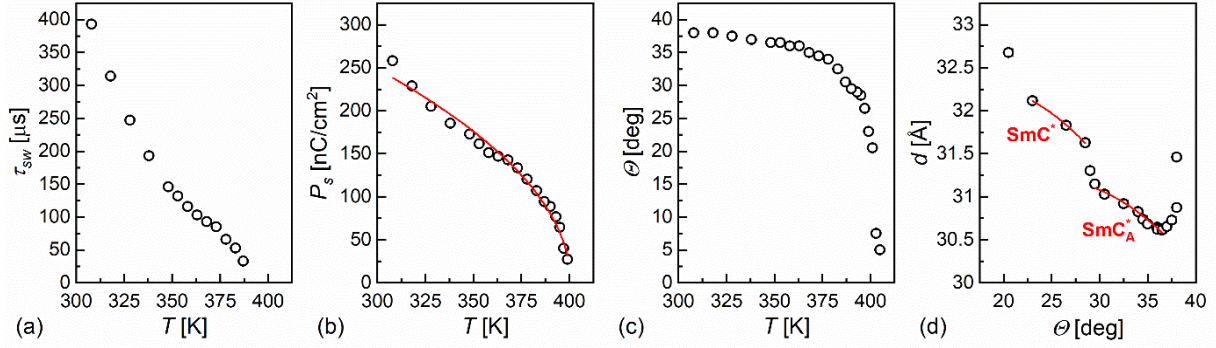

Figure 12. Switching time (a), spontaneous polarization (b), and tilt angle (c) in MIX23HH6 as a function of temperature. Panel (d) shows the smectic layer spacing as a function of the tilt angle.

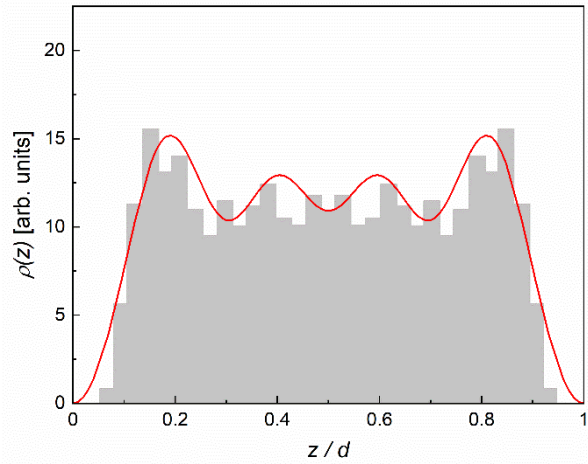

Figure 13. Electron density profile along the smectic layer normal in MIX23HH6, obtained from DFT calculations at the B3LYP-D3(BJ)/6-31+Gd level (average over four conformations and weighted by the molar percentage of each component). Line is a fitting result of Equation (2).

## 4. Summary and conclusions

This study demonstrates that a multicomponent chiral liquid crystalline system can form a vitrified hexatic smectic phase exhibiting unique dielectric and optical properties. The tertiary mixture MIX23HH6, consisting of MHPOBC, 3F2HPhH6, 3F3HPhH6 in molar fractions 0.5, 0.25, 0.25, forms four liquid crystal smectic phases: A*, C*, $C_A$*, and hexatic $X_A$*, which undergoes the glass transition around 270 K. The $SmC_A$* phase is characterized by switching time less than 150 μs down to 348 K and a tilt angle of 35-38° below 368 K. By comparing the temperature dependence of the tilt angle and layer spacing, it is inferred that MHPOBC dimers, which form in a pure compound, are present also in SmC*, but absent in the $SmC_A$* phase of MIX23HH6. Assuming the absence of dimers, an electron density profile perpendicular to smectic layers is obtained from molecular models optimized by DFT method, which is in qualitative agreement with the experimental density profile determined from intensities of diffraction peaks deep in $SmC_A$*.

SRL is observed in all tilted smectic phases of MIX23HH6. SmC* shows SRL of mainly blue light. SmC$_A$* shows SRL with a wavelength decreasing on cooling, from 826 nm (red light) at 326 K to 375 nm (blue light) at 308 K; the last point is already a region of the SmC$_A$*/SmX$_A$* transition. The range of SRL in SmC$_A$* is shifted to higher temperatures by ca. 20-25 K on heating. SmX$_A$* gives a dark blue image under microscope in a reflection mode, suggesting that SRL occurs for UV light, just below the visible region. A change to dark red image occurs shortly before the glass transition of SmX$_A$*, which remains an open question. Another unexpected observation is an apparent absence of the α-process in SmX$_A$*, which is currently explained by overlapping with the β process.

Additional measurements for pure components 3F2HPhH6 and 3F3HPhH6 indicate formation of SmX$_A$*, although it occurs either during crystallization or in a partially crystallized sample and the SmX$_A$* glass is not obtained even at the highest cooling rate of 40 K/min. Relaxation processes are observed for both compounds in a crystalline state. The corresponding energy barriers of 30-70 kJ/mol suggest an origin in intramolecular rotations and qualify these crystal phases as CONDIS phases. DSC and XRD results for 3F3HPhH6 suggest that another smectic phase may be present in a very narrow temperature range between SmC$_A$* and isotropic liquid, but it could not have been identified.

Future work will focus on additional 3FmHPhH6/MHPOBC systems, particularly observation of α- and β-relaxation processes in SmX$_A$*, and search of mixtures showing SRL in visible range in SmX$_A$*, which will facilitate study of a helical ordering in this phase.

**Acknowledgements:** We thank Dr. Eng. Magdalena Urbańska from the Institute of Chemistry of the Military University of Technology in Warsaw for chemical synthesis. We gratefully acknowledge Polish high-performance computing infrastructure PLGrid (HPC Center: ACK Cyfronet AGH) for providing computer facilities and support within computational grant no. PLG/2026/019109. Michał Czerwiński and Mateusz Filipow thank the MUT University grant UGB 22-094 for the financial support. Aleksandra Deptuch acknowledges the National Science Centre, Poland (grant MINIATURA 7 no. 2023/07/X/ST3/00182) for financial support.

**Conflicts of interest statement:** There are no conflicts to declare.

**Authors' contributions:**
A. Deptuch – conceptualization, investigation, formal analysis, funding acquisition, writing – original draft
A. Drzewicz – investigation, writing – review and editing
M. Piwowarczyk – investigation, writing – review and editing
M. Czerwiński – investigation, writing – review and editing
M. Filipow – investigation, writing – review and editing
M. Pączek – investigation, writing – review and editing
E. Juszyńska-Gałązka – investigation, writing – review and editing

# Ternary liquid crystalline mixture showing broad antiferroelectric smectic C$_A$* and glassy hexatic smectic X$_A$* phases

Aleksandra Deptuch[1,*], Anna Drzewicz[1], Marcin Piwowarczyk[1], Michał Czerwiński[2], Mateusz Filipow[2], Mateusz Pączek[3], Ewa Juszyńska-Gałązka[1,4]

[1] Institute of Nuclear Physics, Polish Academy of Sciences, Radzikowskiego 152, PL-31342 Kraków, Poland

[2] Institute of Chemistry, Military University of Technology, Kaliskiego 2, PL-00908 Warsaw, Poland

[3] Faculty of Chemistry, Jagiellonian University, Gronostajowa 2, PL-30387 Kraków, Poland

[4] Research Center for Thermal and Entropic Science, Graduate School of Science, Osaka University, 560-0043 Osaka, Japan

*corresponding author, aleksandra.deptuch@ifj.edu.pl

# Supplementary Materials

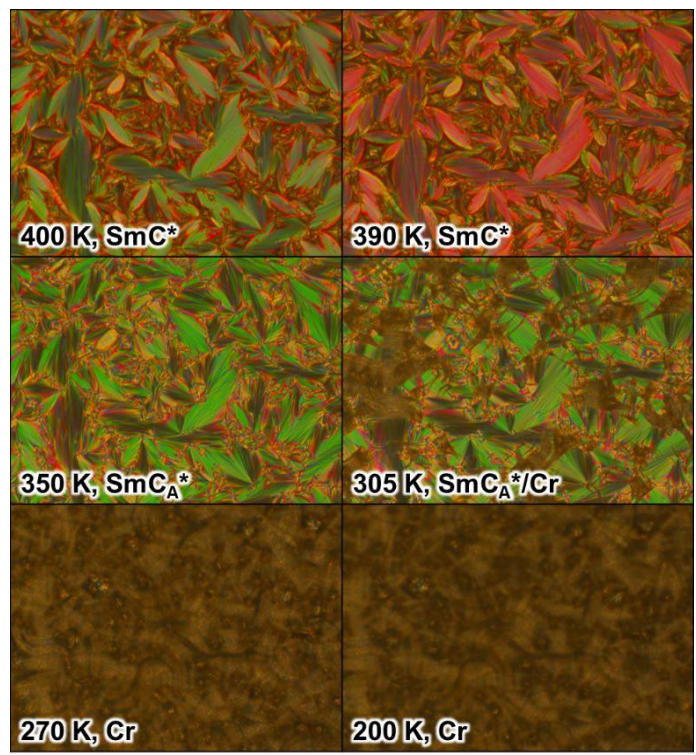

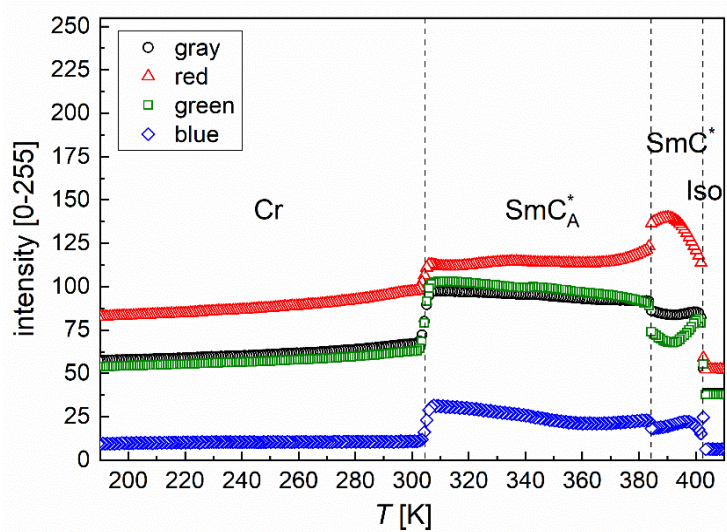

Figure S1. POM textures (622 × 466 μm$^2$) of 3F2HPhH6 collected at the 10 K/min cooling rate in the transmission mode. The plot below shows the red, green, blue components and weighted total intensity of each texture.

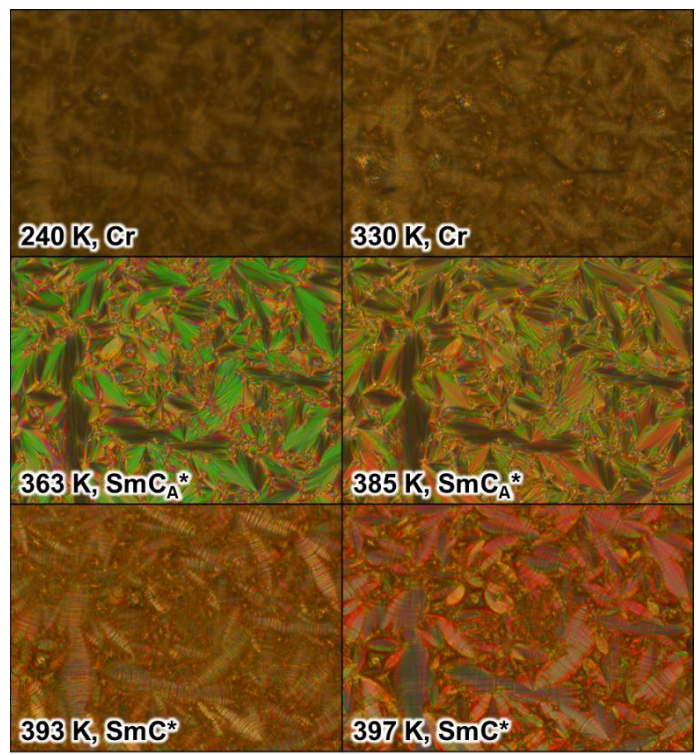

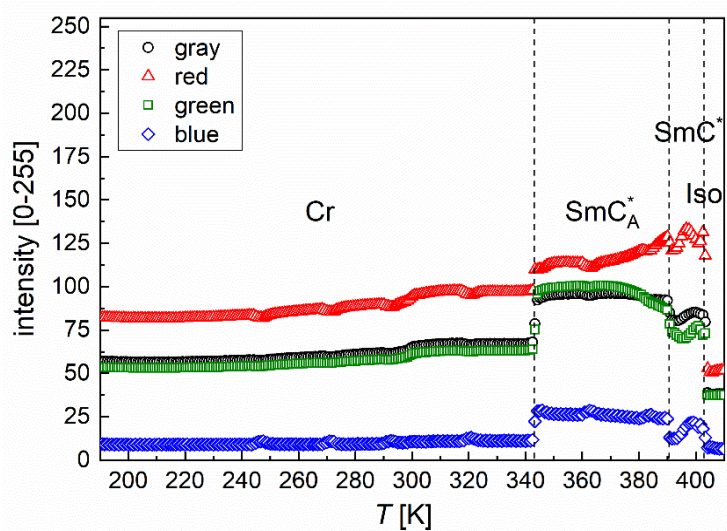

Figure S2. POM textures (622 × 466 μm$^2$) of 3F2HPhH6 collected at the 10 K/min heating rate in the transmission mode. The plot below shows the red, green, blue components and weighted total intensity of each texture.

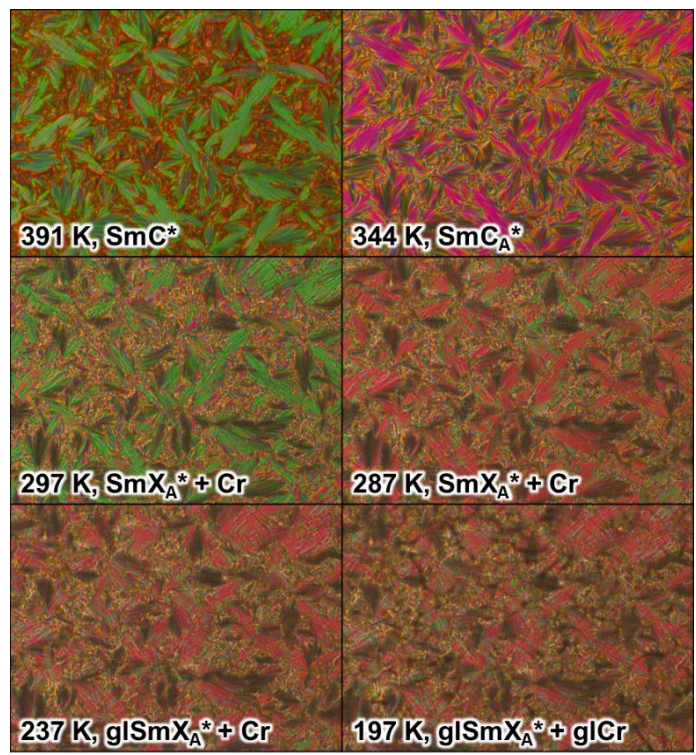
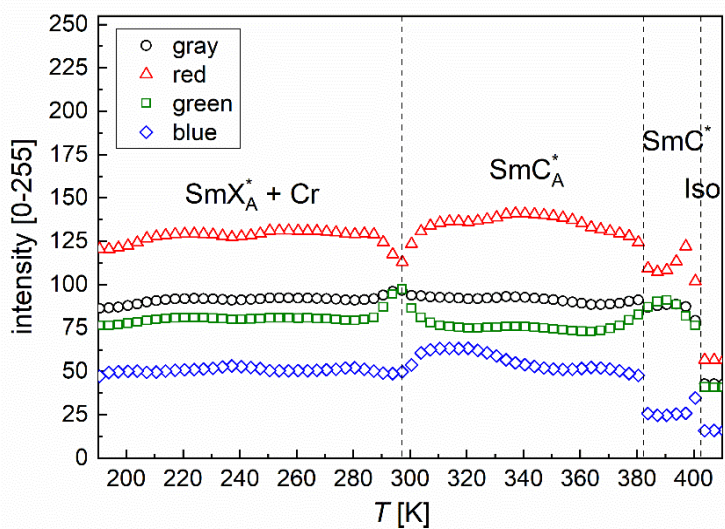

Figure S3. POM textures (622 × 466 μm$^2$) of 3F2HPhH6 collected at the 40 K/min cooling rate in the transmission mode. The plot below shows the red, green, blue components and weighted total intensity of each texture.

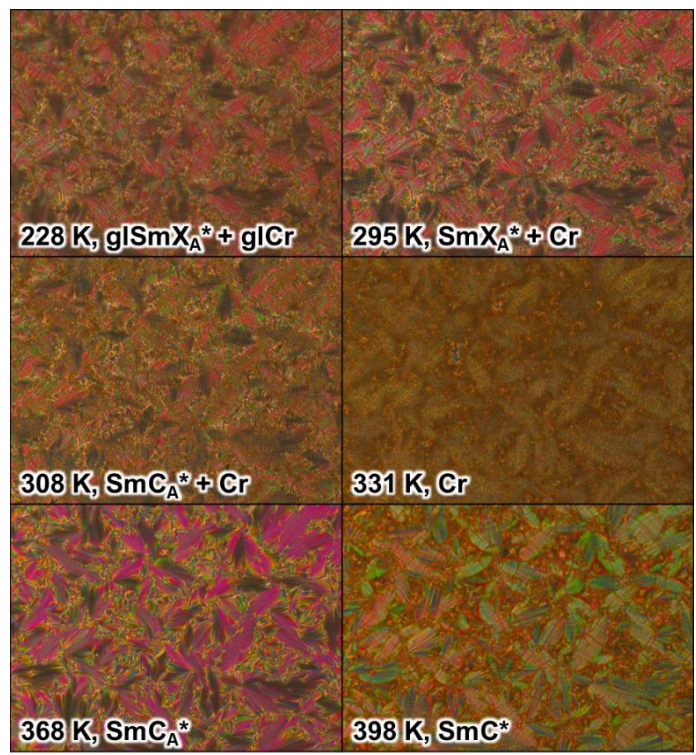
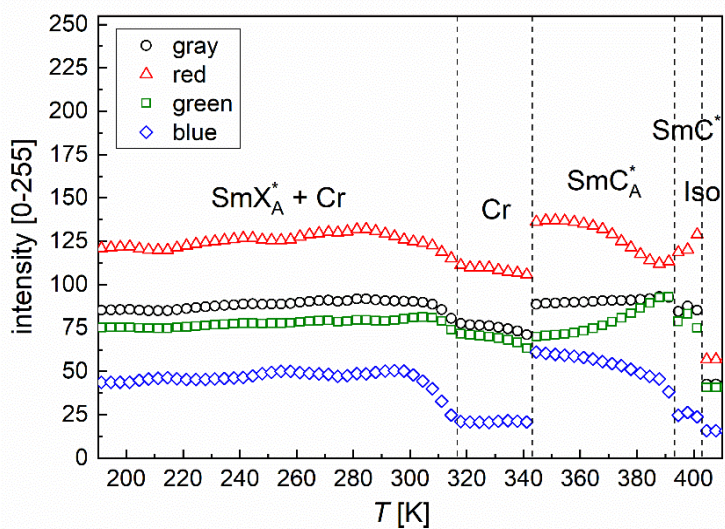

Figure S4. POM textures (622 × 466 μm$^2$) of 3F2HPhH6 collected at the 40 K/min heating rate in the transmission mode. The plot below shows the red, green, blue components and weighted total intensity of each texture.

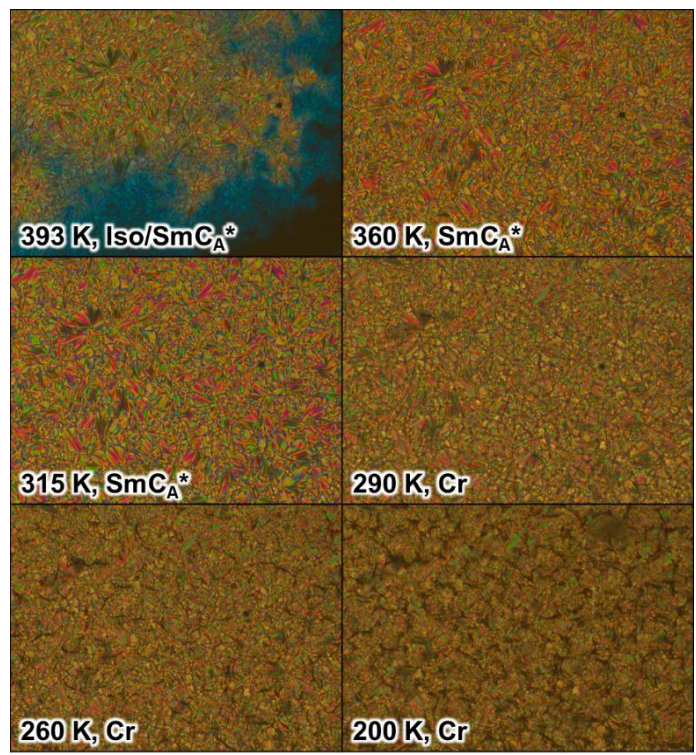

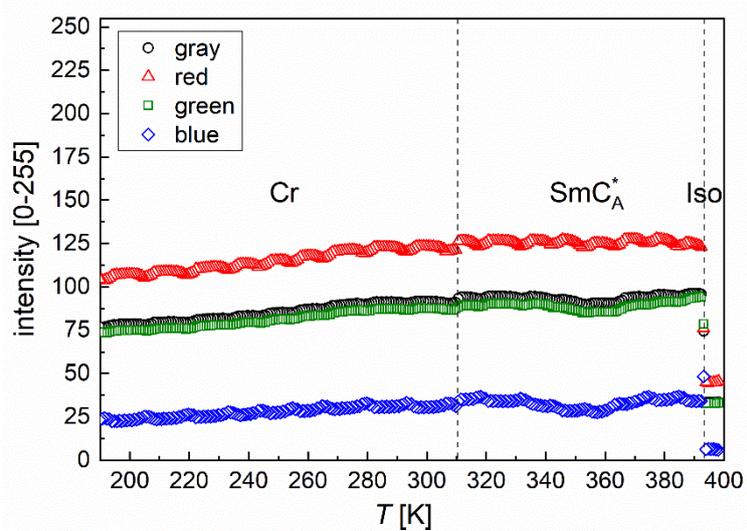

Figure S5. POM textures (622 × 466 μm$^2$) of 3F3HPhH6 collected at the 10 K/min cooling rate in the transmission mode. The plot below shows the red, green, blue components and weighted total intensity of each texture.

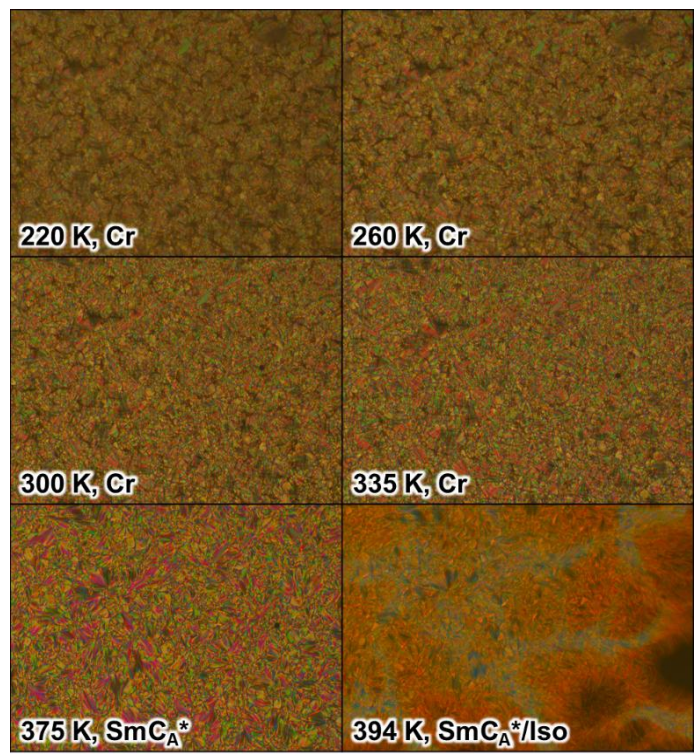
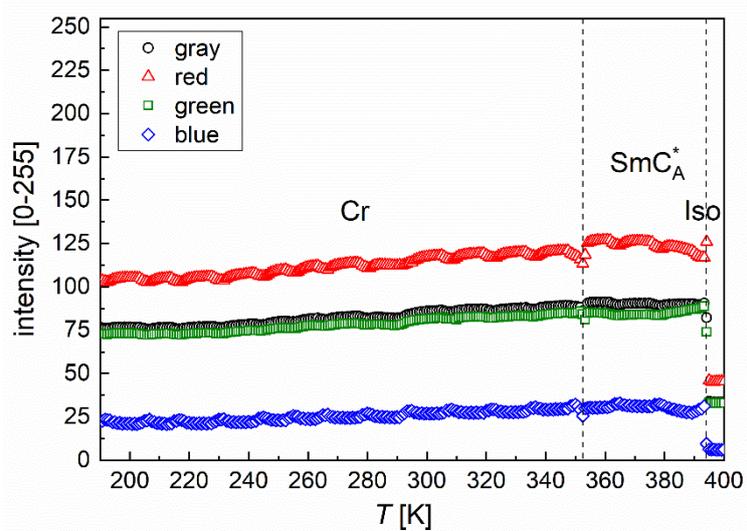

Figure S6. POM textures (622 × 466 μm$^2$) of 3F3HPhH6 collected at the 10 K/min heating rate in the transmission mode. The plot below shows the red, green, blue components and weighted total intensity of each texture.

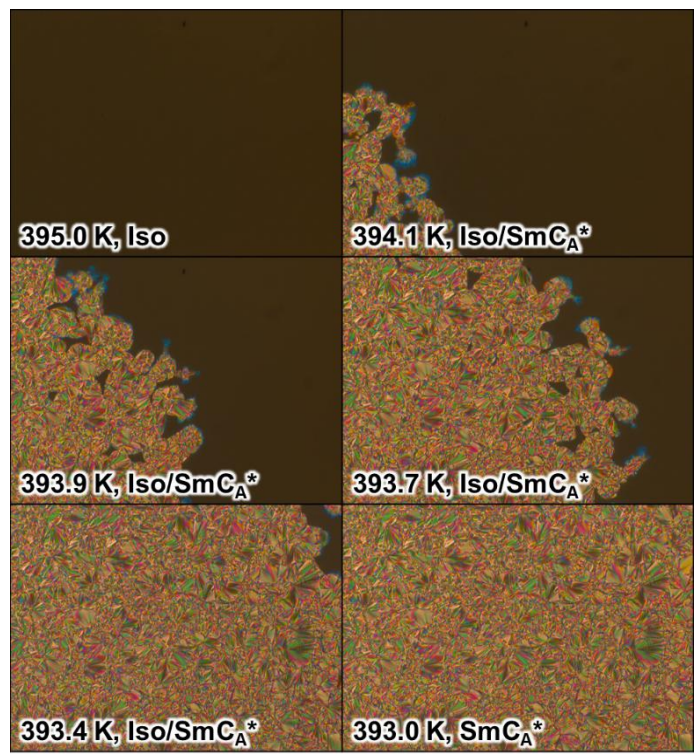

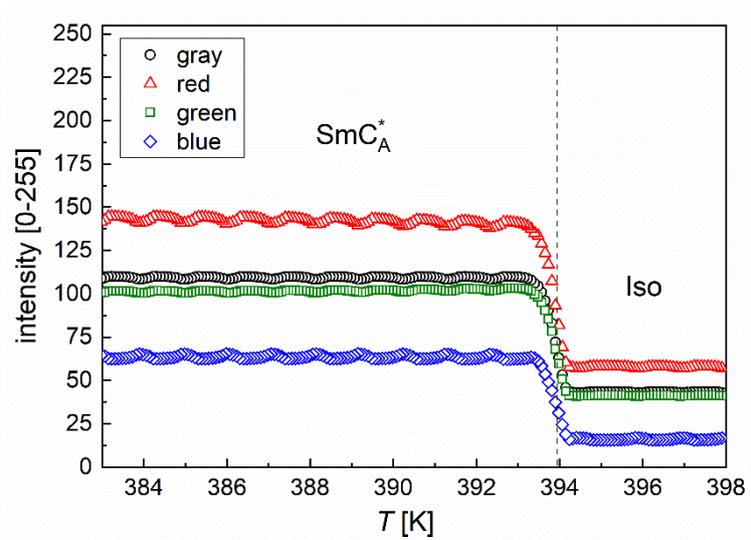

Figure S7. POM textures (622 × 466 μm$^2$) of 3F3HPhH6 collected at the 1 K/min cooling rate in the transmission mode. The plot below shows the red, green, blue components and weighted total intensity of each texture.

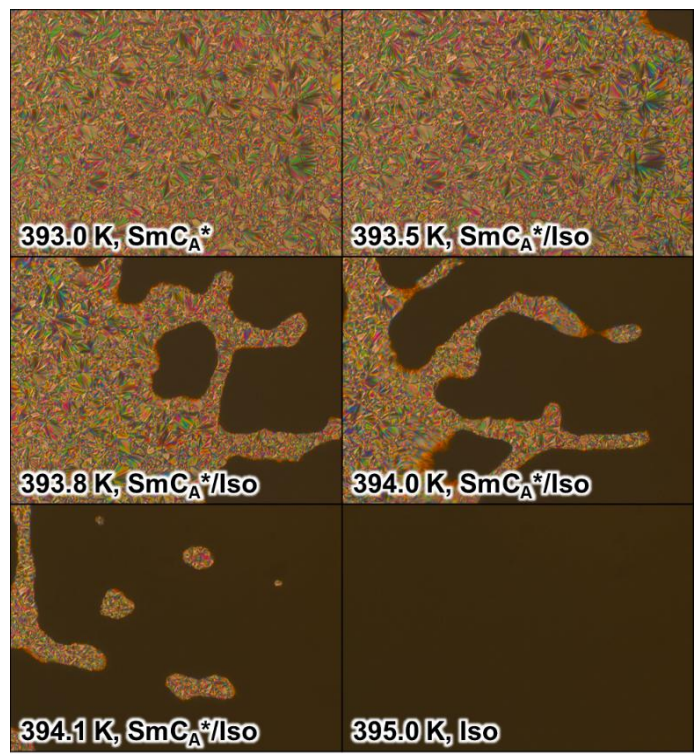
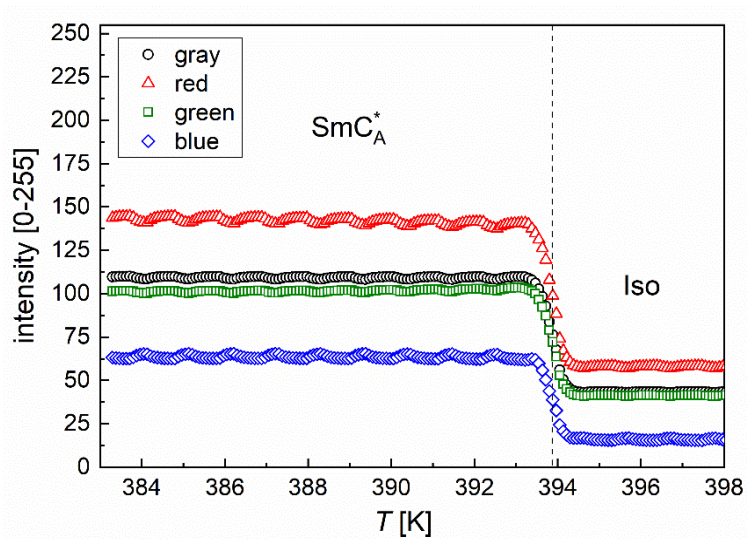

Figure S8. POM textures (622 × 466 μm$^2$) of 3F3HPhH6 collected at the 1 K/min heating rate in the transmission mode. The plot below shows the red, green, blue components and weighted total intensity of each texture.

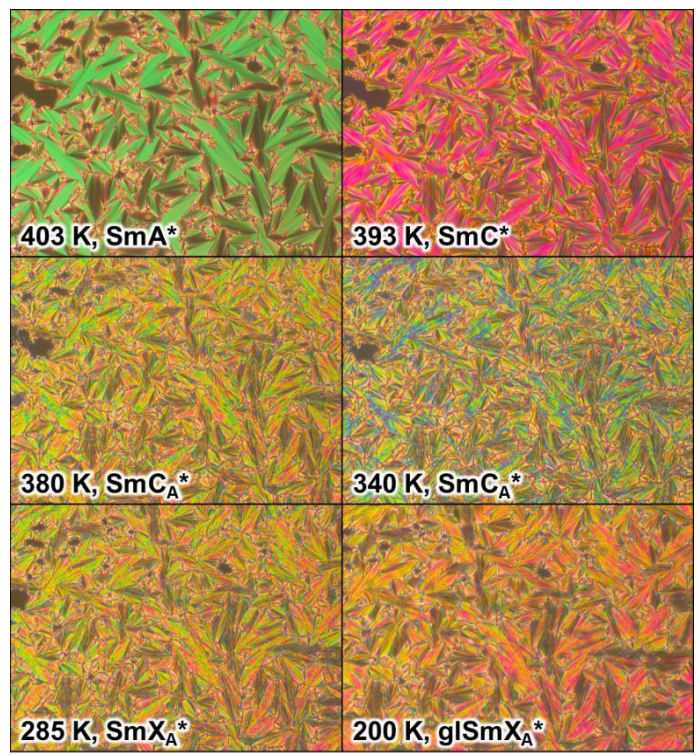

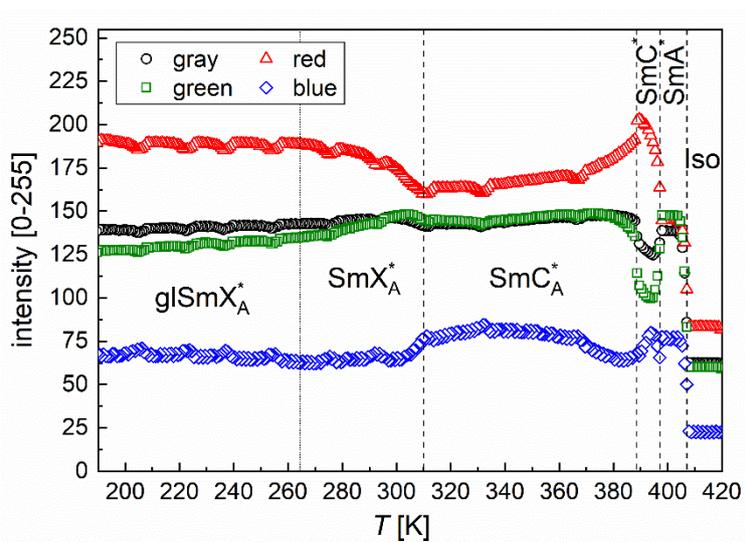

Figure S9. POM textures (622 × 466 μm$^2$) of MIX23HH6 collected at the 10 K/min cooling rate in the transmission mode. The plot below shows the red, green, blue components and weighted total intensity of each texture. The glass transition temperature is based on the corresponding DSC results.

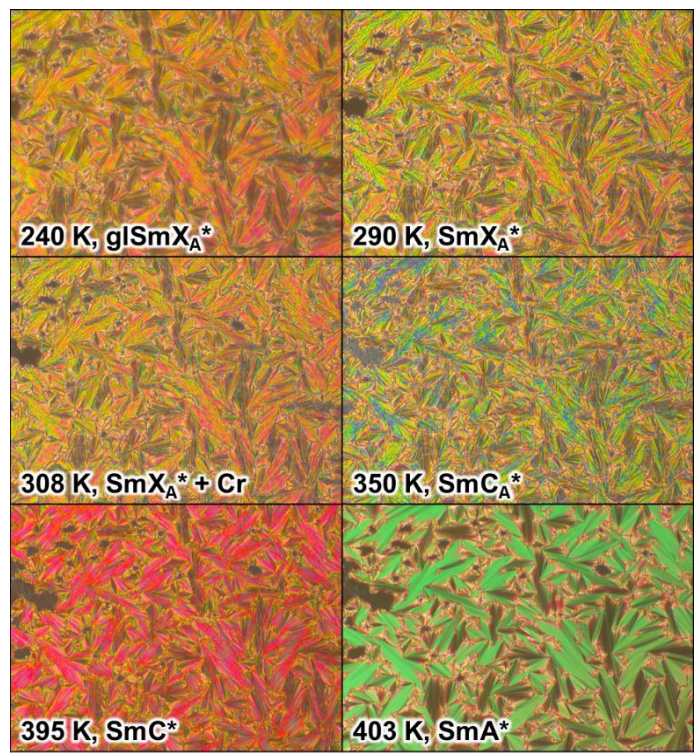
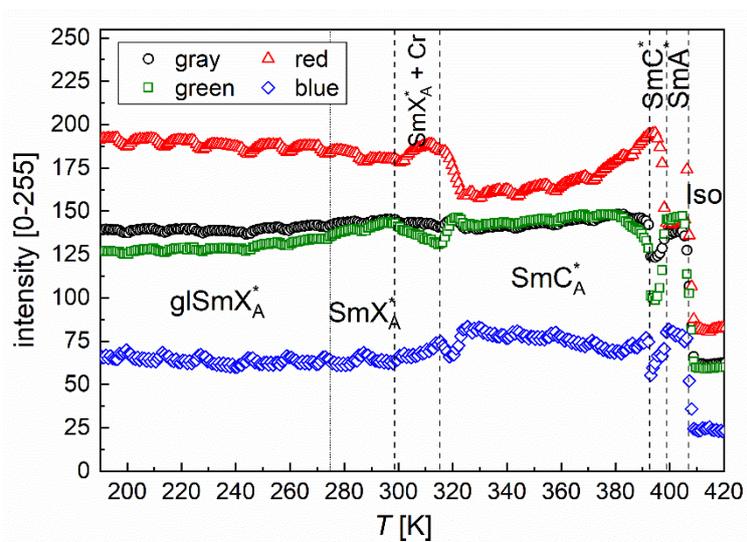

Figure S10. POM textures (622 × 466 μm$^2$) of MIX23HH6 collected at the 10 K/min heating rate in the transmission mode. The plot below shows the red, green, blue components and weighted total intensity of each texture. The glass transition temperature is based on the corresponding DSC results.

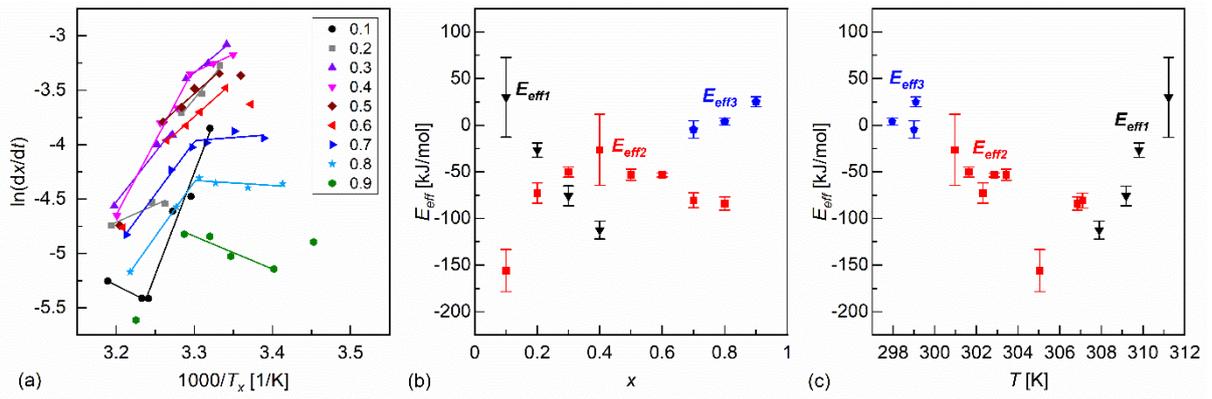

Figure S11. The melt crystallization of 3F2HPhH6 analyzed by the isoconversional method: activation plot (a), effective activation energy vs. conversion degree (b), effective activation energy vs. average temperature.

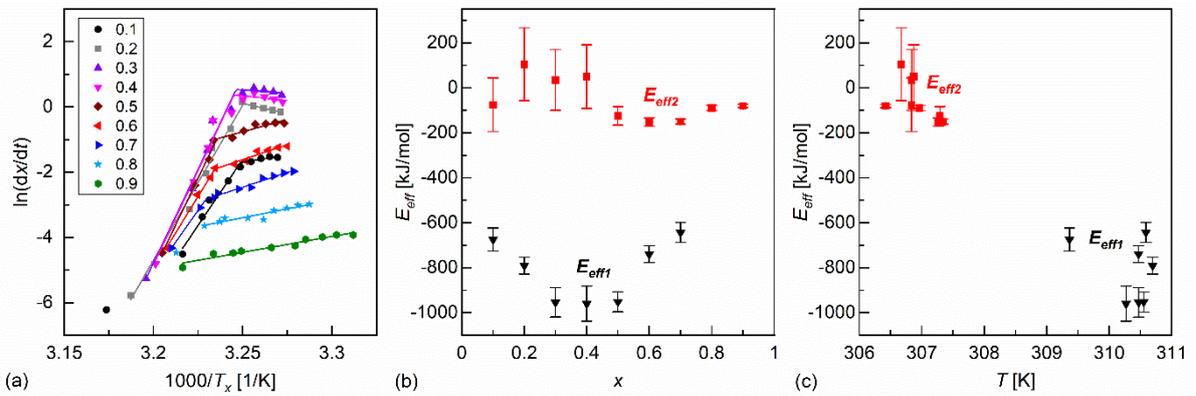

Figure S12. The melt crystallization of 3F3HPhH6 analyzed by the isoconversional method: activation plot (a), effective activation energy vs. conversion degree (b), effective activation energy vs. average temperature.

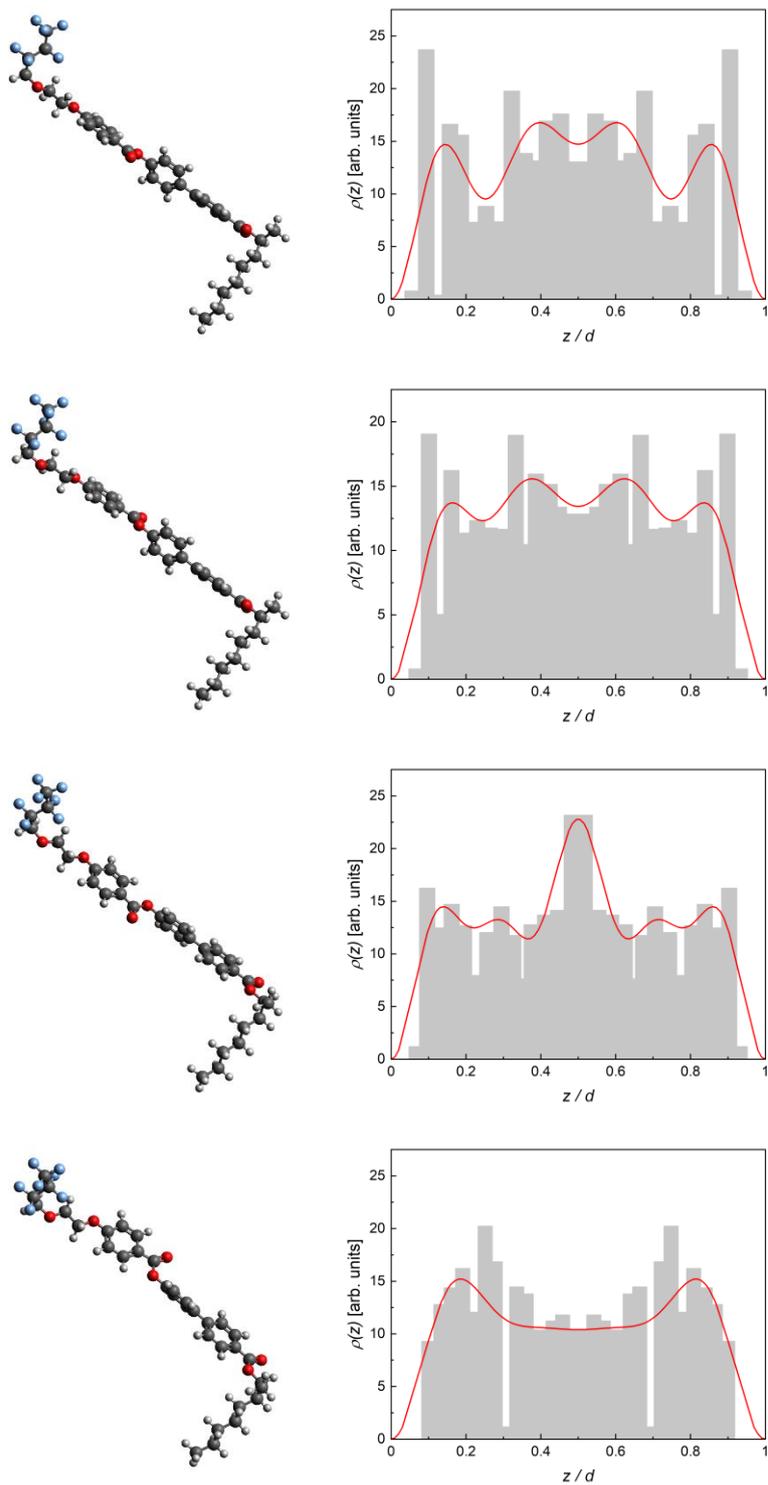

Figure S13. Models of 3F2HPhH6 optimized at the B3LYP-D3(BJ)/6-31+Gd level and corresponding electron density profiles along the smectic layer normal. Lines are fitting results of Equation (2) from the main text.

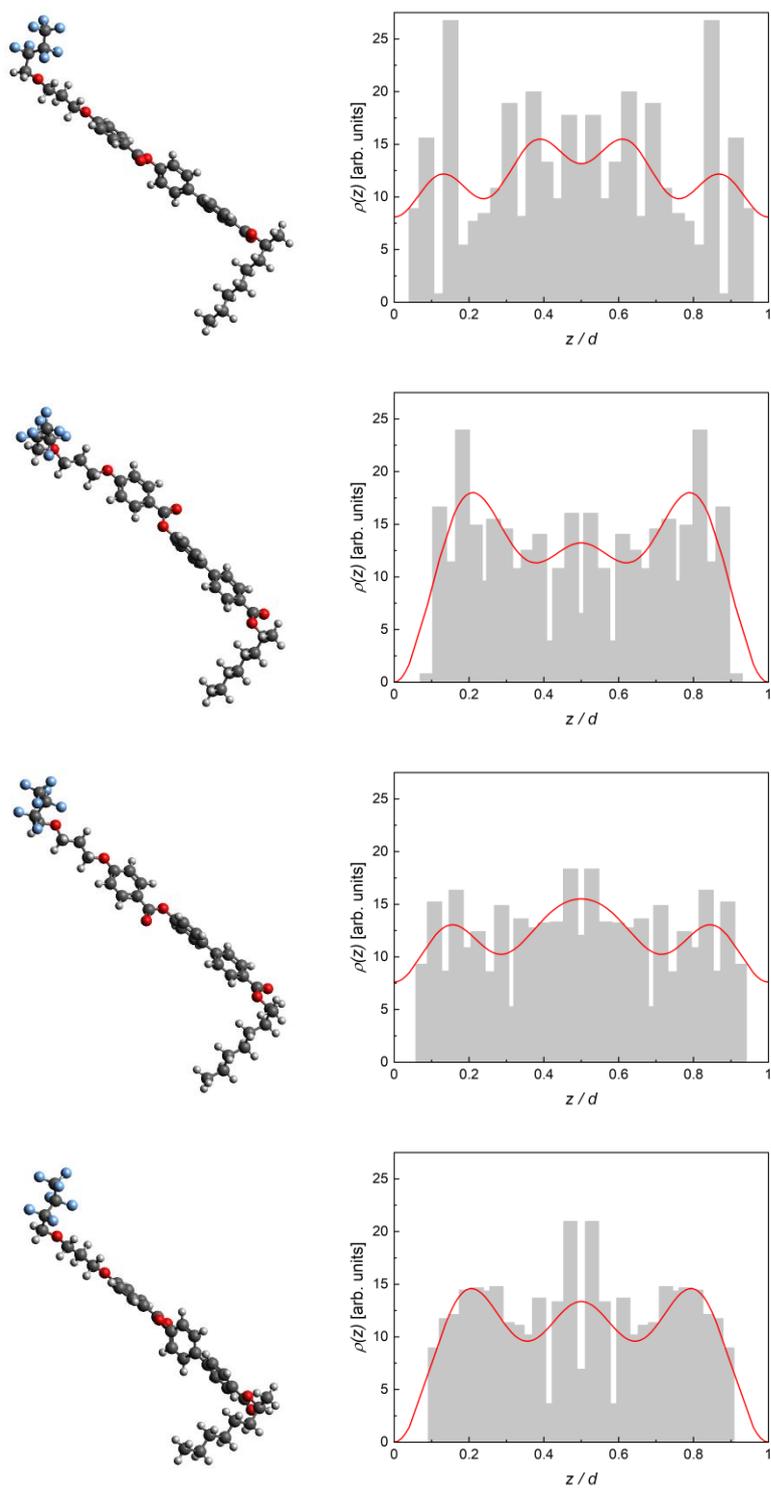

Figure S14. Models of 3F3HPhH6 optimized at the B3LYP-D3(BJ)/6-31+Gd level and corresponding electron density profiles along the smectic layer normal. Lines are fitting results of Equation (2) from the main text.

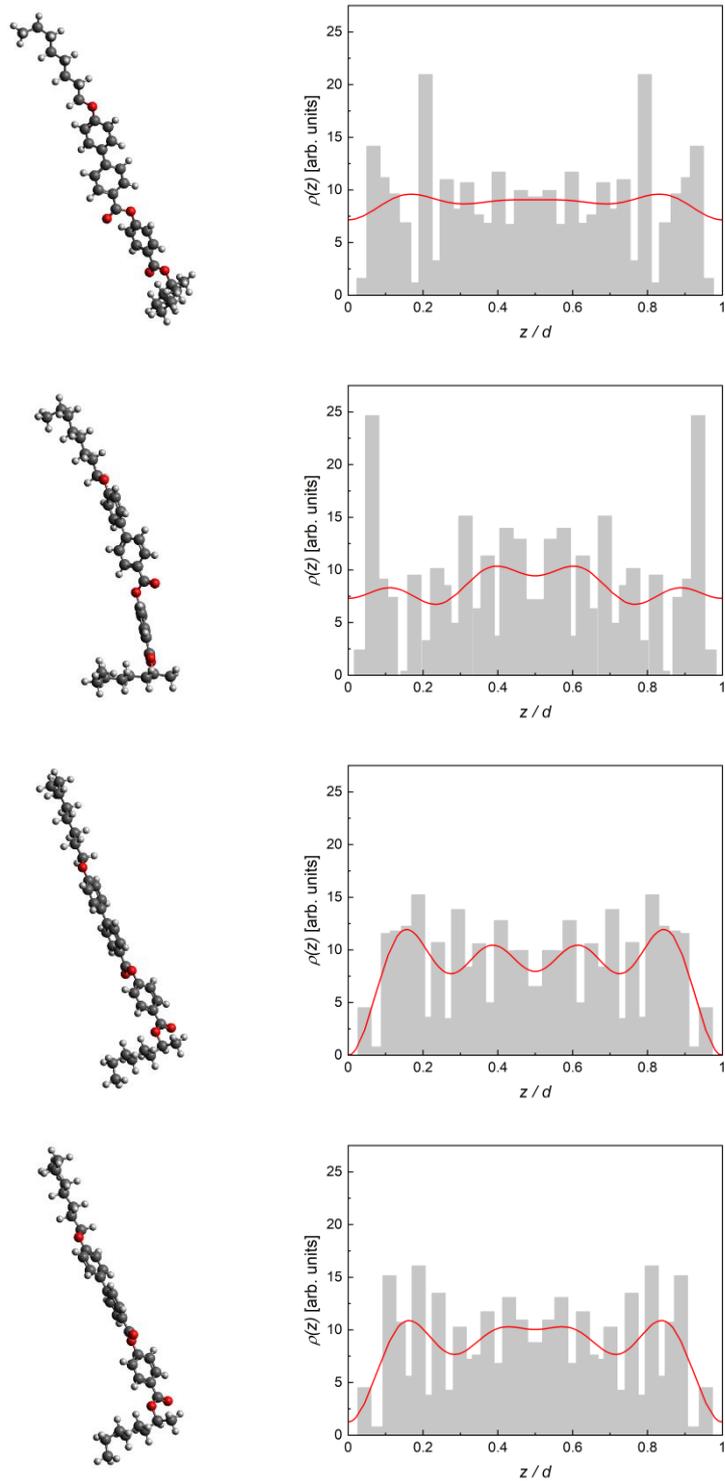

Figure S15. Models of MHPOBC optimized at the B3LYP-D3(BJ)/6-31+Gd level and corresponding electron density profiles along the smectic layer normal. Lines are fitting results of Equation (2) from the main text.

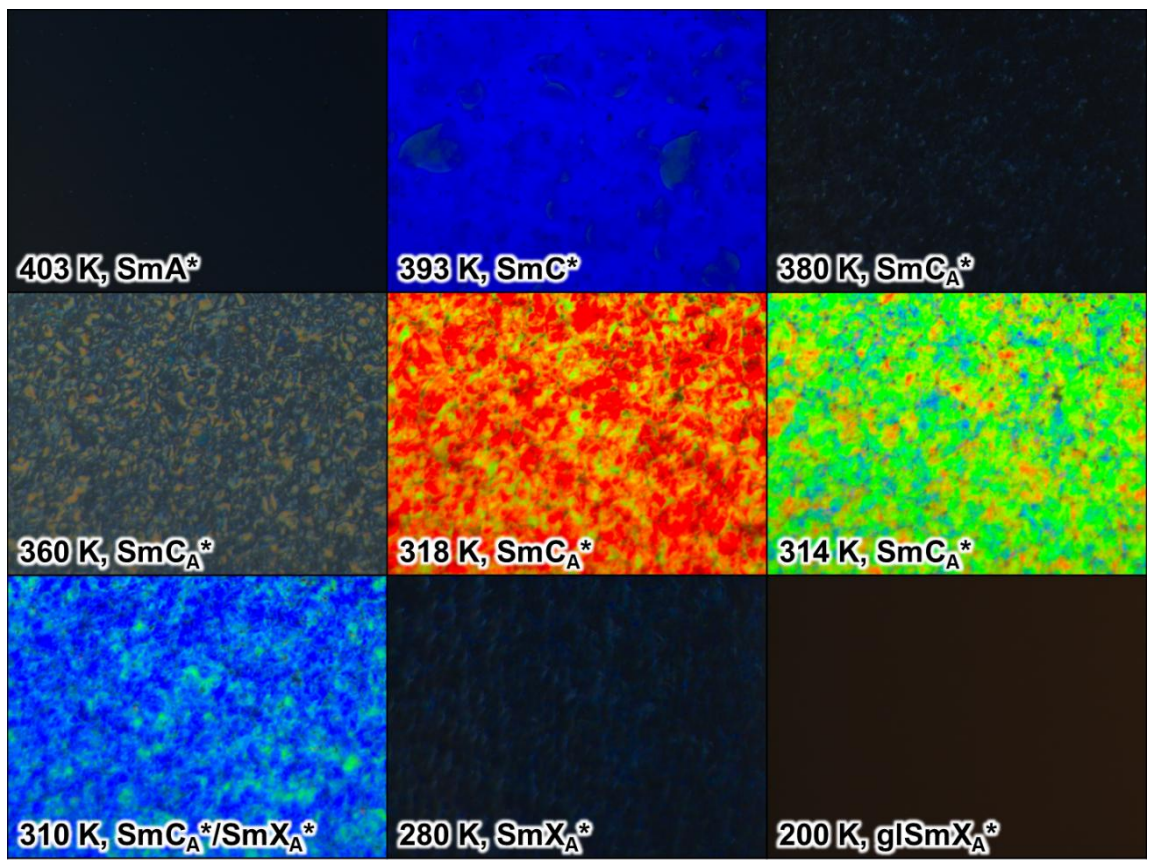

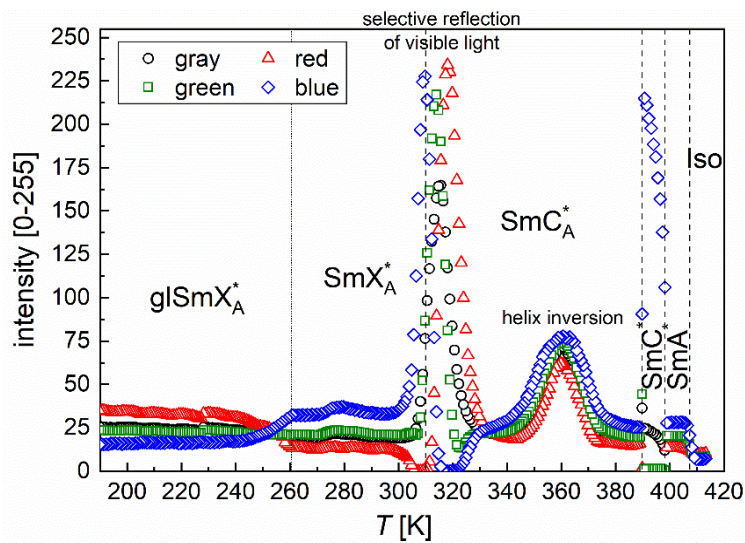

Figure S16. POM textures (622 × 466 μm$^2$) of MIX23HH6 collected at the 10 K/min cooling rate in the reflection mode. The plot below shows the red, green, blue components and weighted total intensity of each texture. The SmC$_A$* → SmX$_A$* transition temperature is based on the data from Figure S9.

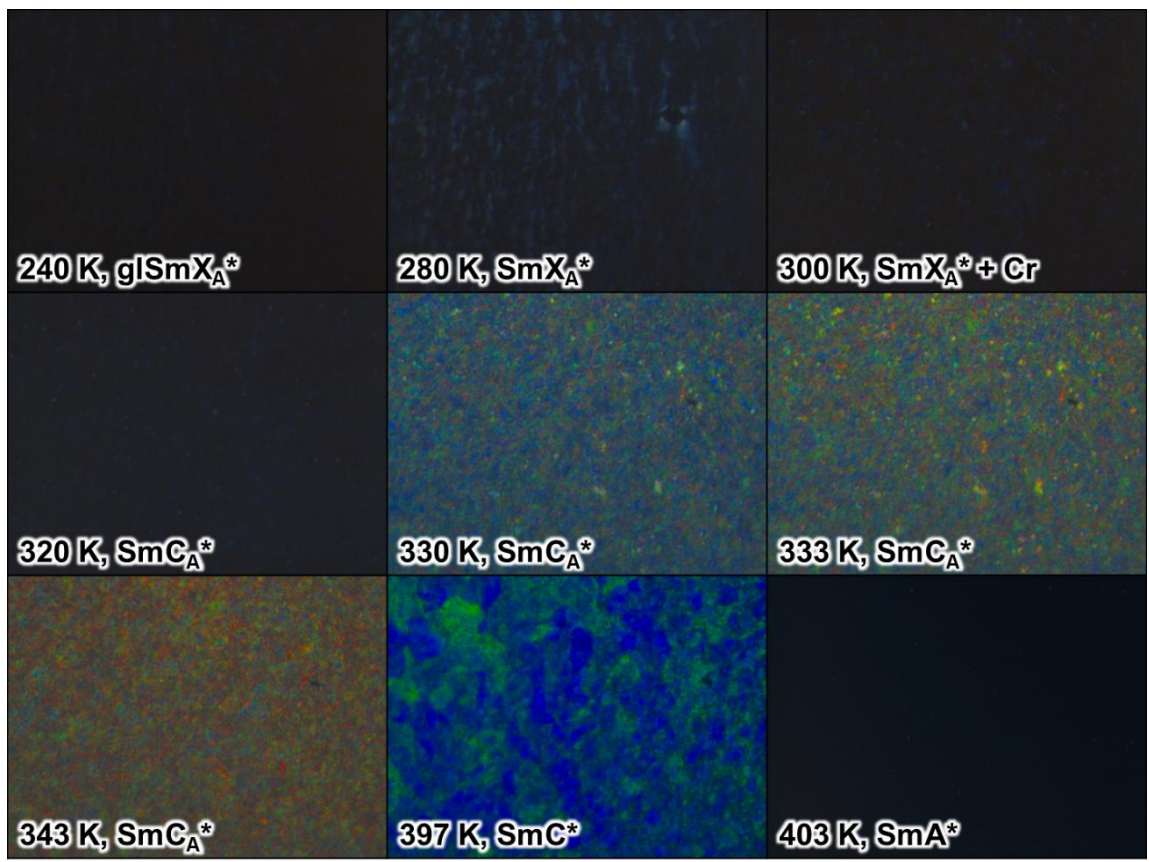

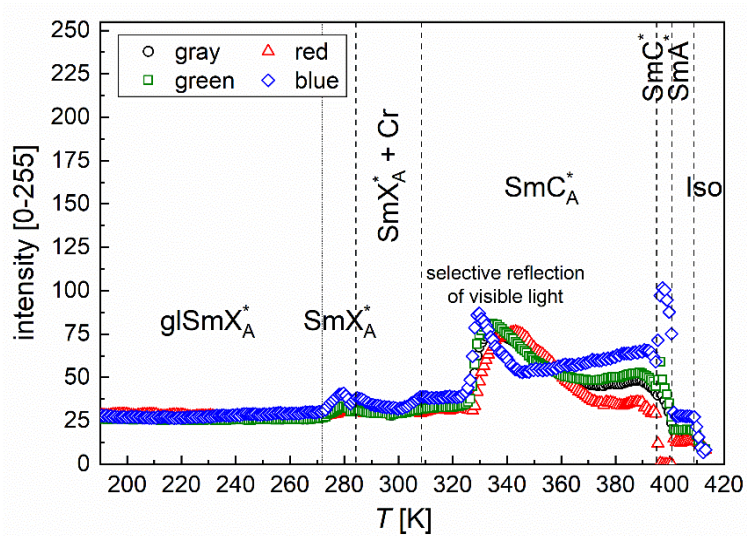

Figure S17. POM textures (622 × 466 μm$^2$) of MIX23HH6 collected at the 10 K/min heating rate in the reflection mode. The plot below shows the red, green, blue components and weighted total intensity of each texture.

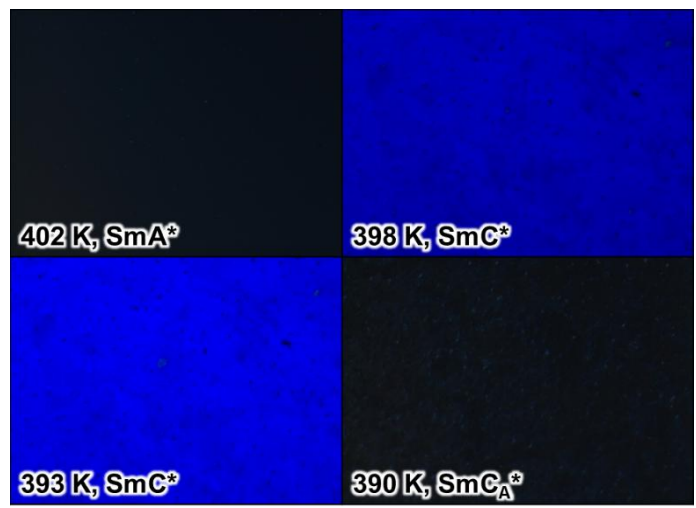

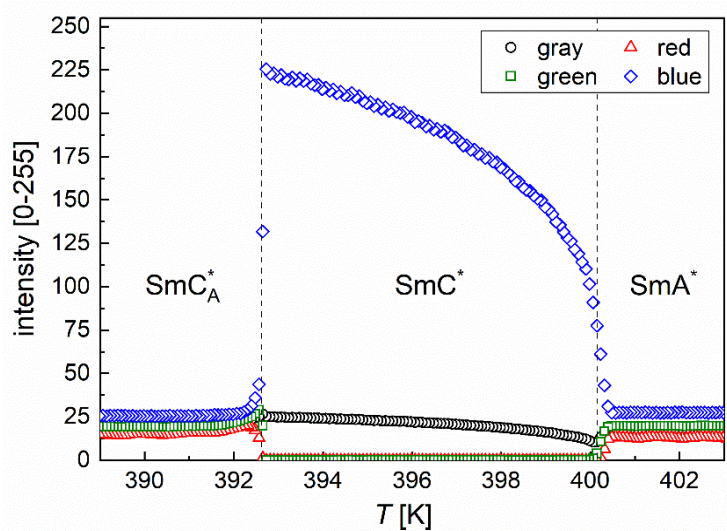

Figure S18. POM textures (622 × 466 μm$^2$) of MIX23HH6 collected at the 1 K/min cooling rate in the reflection mode. The plot below shows the red, green, blue components and weighted total intensity of each texture.

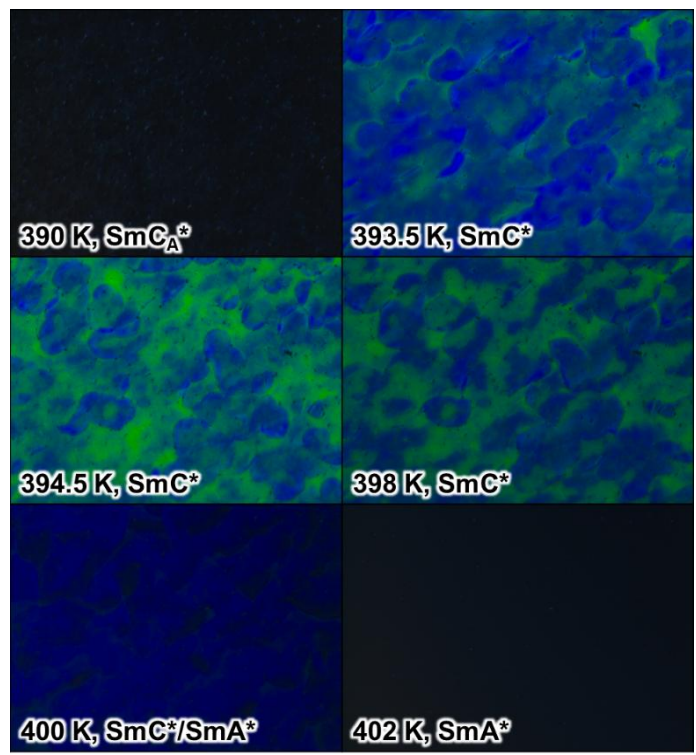

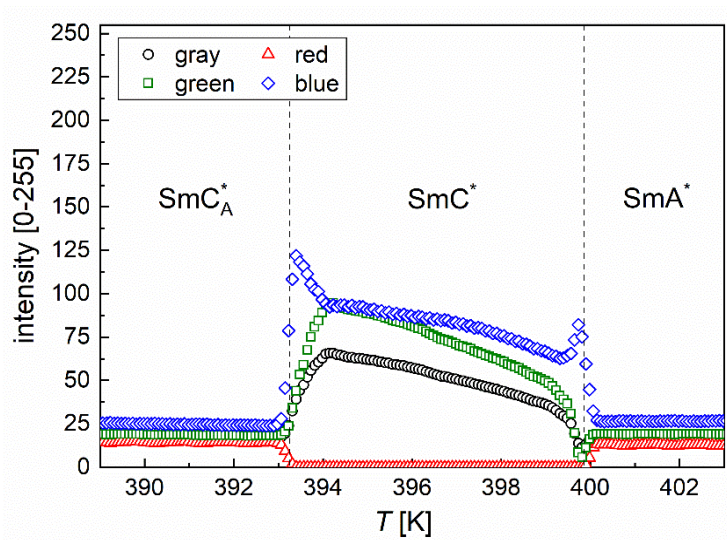

Figure S19. POM textures (622 × 466 μm$^2$) of MIX23HH6 collected at the 1 K/min heating rate in the reflection mode. The plot below shows the red, green, blue components and weighted total intensity of each texture.

41